\documentclass[12pt]{article}
\usepackage{amsmath, amsthm, amssymb}
\usepackage{achemso}
\usepackage{epsfig}
\begin{document}
\baselineskip= 24pt
\begin{center}
{\large Adsorption of Multi-block and Random Copolymer on a Solid Surface:\\
Critical Behavior and Phase Diagram} \\
\vskip 0.2 true cm
{S. Bhattacharya$^1$, H.-P. Hsu$^2$, A. Milchev$^{1,3}$,
V. G. Rostiashvili$^1$ and T. A. Vilgis$^1$}
\vskip 0.2true cm
{$^1$ Max Planck Institute for Polymer Research
10 Ackermannweg, 55128 Mainz, Germany\\
$^2$ Institute of Physics, Johannes Gutenberg-University,
Staudinger Weg 7, 55099 Mainz, Germany \\
$^3$ Institute for Physical Chemistry, Bulgarian Academy of Science,
1113 Sofia, Bulgaria}
\end{center}
\date{\today}

\begin{abstract}
The adsorption of a single multi-block $AB$-copolymer on a solid planar
substrate is investigated by means of computer simulations and scaling analysis.
It is shown that the problem can be mapped onto an effective homopolymer
adsorption problem. In particular we discuss how the critical adsorption energy
and the fraction of adsorbed monomers depend on the block length $M$ of sticking
monomers $A$, and on the total length $N$ of the polymer chains. Also the
adsorption of the random copolymers is considered and found to be well
described within the framework of the annealed approximation. For a better test
of our theoretical prediction, two different Monte Carlo (MC) simulation methods
were employed: a) off-lattice dynamic bead-spring model, based on the
standard Metropolis algorithm (MA), and b) coarse-grained lattice model using
the Pruned-enriched Rosenbluth method (PERM) which enables tests for very long
chains. The findings of both methods are fully consistent and in good agreement
with theoretical predictions.

\end{abstract}



\section{Introduction}
Adsorption of polymers on surfaces plays a key role in numerous technological
applications and is also relevant to many biological processes. During the last
three decades it has been constantly a focus of research interest. The
theoretical studies of the behavior of polymers interacting with solid substrate
have been based predominantly on both scaling
analysis~\cite{Gennes_1,Gennes_2,Gennes_3,Gennes_4,Eisen} as well as on the
self-consistent field (SCF) approach~\cite{Fleer}. The close relationship
between theory and computer experiments in this field\cite{Eisen,AMKB} has
proved especially fruitful. Most investigations focus as a rule on the
determination of the critical adsorption point (CAP) location and on the scaling
behavior of a variety of quantities below, above and at the CAP. Thus an eminent
relation between polymer statistics and the corresponding correlation
functions~\cite{Eisen} in the $n$-vector model of magnets with a free surface in
the limit $n \rightarrow 0$ has lead to a number of important results.
Special interest has been payed to the determination of the so called
{\em crossover exponent} $\phi$ which is known to govern the fraction of
adsorbed monomers at the CAP. Recently the scaling relationship for a single
chain adsorption has been tested by Monte Carlo (MC) simulation on a cubic
lattice~\cite{Sommer,Grassberger} as well as by an off-lattice
model~\cite{Metzger,AMKB} and the adsorption transition of a polymer could be
viewed
nowadays as comparatively well understood.

While the investigations mentioned above have been devoted exclusively to
homopolymers, the adsorption of copolymers (e.g. multi-blocks or random
copolymers) is still much less understood. Thus, for instance, the CAP
dependence on block size $M$ at fixed concentration of the sticking $A$-mers is
still unknown as are the scaling properties of {\em regular multi-block
copolymers} in the vicinity of the CAP. From the theoretical perspective, the
case of diblock copolymers has been studied mainly within the
SCF-approach~\cite{Fleer,Evers}. The case of {\em random copolymers} adsorption
has gained comparatively more attention by researcher so far. It has been
investigated by Whittington et al.~\cite{Whit_1,Whit_2} using both the annealed
and quenched models of randomness. In the latter case the authors implemented
the Morita approximation (which is reduced to an optimization problem with a set
of constraints involving the moments of the quenched random probability
distribution). The influence of sequence correlations on the adsorption of
random copolymers has been studied by means of the variational and replica
method approach\cite{Polotsky}. Sumithra and Baumgaertner~\cite{Baum} examined
the question of how the critical behavior of random copolymers differs from that
of homopolymers. Thus, among a number of important conclusions, the results of
Monte Carlo simulations demonstrated that the crossover exponent $\phi$ (see
below) is independent of the fraction of attractive monomers $f$.

In the present paper we use scaling analysis as well as two MC-simulation
methods to study the critical behavior of multi-block and random copolymers. It
turns out that the critical behaviour of these two types of copolymers could be
reduced to the behavior of an effective homopolymer chain with "renormalized"
segments. For the multi-block copolymer this allows e.g. to explain how the
critical attraction energy depends on the block length $M$ and to derive an
adsorption phase diagram in terms of CAP against $M$. In the case of random
copolymers the sequence of sticky and neutral (as regards the solid substrate)
monomers within a particular chain is fixed which exemplifies a system with
quenched randomness. Nevertheless, close to criticality the chain is still
rather mobile, so that the sequence dependence is effectively averaged over the
time of the experiment and the problem can be reduced to the case of annealed
randomness. We show that our MC-findings close to criticality could be perfectly
treated within the annealed randomness model.

\section{Scaling properties of homopolymer adsorption}
\label{Theory}
\subsection{Order parameter}\label{OP}

Before discussing copolymers adsorption we briefly sketch the scaling theory of
homopolymer adsorption~\cite{Eisen,Sommer,Metzger}. It is well known that a
single polymer chain undergoes a transition from a non-bound into an adsorbed
state when the adsorption energy $\epsilon$ per monomer increases beyond a
critical value $\epsilon_c$. Here and in what follows $\epsilon$ is measured in
units of the thermal energy $k_BT$ (with $k_B$ being the Boltzmann constant,
ane $T$ - the temperature of the system). The adsorption transition can be
interpreted as a second-order phase transition at the critical point (CAP) of
adsorption $\epsilon = \epsilon_c$ in the thermodynamical limit, i.e. $N
\rightarrow \infty$. Close to the CAP the number of surface contacts $N_s$
scales as $N_s (\epsilon = \epsilon_c)\sim N^{\phi}$. The numerical value of
$\phi$ is somewhat controversial and lies in a range between $\phi = 0.59$ (ref.
\cite{Eisen}) and $\phi = 0.484$ (ref.~\cite{Grassberger}), we adopt
however the value $\phi = 0.50\pm 0.02$ which has been suggested as the most
satisfactory\cite{Metzger} by comparison with comprehensive simulation results.

Consider a chain tethered to the surface at the one end. The fraction of
monomers on the surface $f = N_s/N$ may be viewed as an order parameter
measuring the degree of adsorption. In the
thermodynamic limit $N \rightarrow \infty$, the fraction $f$ goes to zero
($\approx {\cal O}(1/N)$) for $\epsilon << \epsilon_c$, then near $\epsilon_c$,
$f \sim N^{\phi - 1}$, and for $\epsilon \gg \epsilon_c$ (in the strong coupling
limit) $f$ it is independent of $N$. Let us measure the distance from the CAP
 by the dimensionless quantity $\kappa = (\epsilon - \epsilon_c)/\epsilon_c$ and
also introduce the scaling variable $\eta \equiv \kappa N^{\phi}$.
The corresponding scaling ansatz is then
\begin{eqnarray}
f(\eta) = N^{\phi - 1} \: G \left(\eta\right)\;.
\label{Scaling_first}
\end{eqnarray}
with the scaling function 
\begin{eqnarray}
G (\eta)=
\begin{cases}
{\rm const} &\quad, \quad {\rm for} \quad \eta \rightarrow 0\\
\eta^{(1 - \phi)/\phi} &\quad, \quad {\rm for } \quad \eta \gg 1
\end{cases}
\label{Scaling_two}
\end{eqnarray}
The resulting scaling behavior of $f$  follows as,
\begin{eqnarray}
f \propto \begin{cases} 1/N &\quad, \quad {\rm for} \quad \kappa << 0\\
N^{\phi - 1} &\quad, \quad {\rm for} \quad  \kappa\rightarrow 0\\
\kappa^{(1 - \phi)/\phi} &\quad, \quad {\rm for} \quad \kappa  \gg 1
\end{cases}
\label{Order_param}
\end{eqnarray}

\subsection{Gyration radius}

The gyration radius in direction perpendicular to the surface,
$R_{g\perp} (\eta)$, has the form
\begin{eqnarray}
R_{g\perp} (\eta) = a N^{\nu} {\cal G}_{g\perp} \left( \eta\right)
\label{Rad_perp}
\end{eqnarray}
One may determine the form of the scaling function ${\cal G}_{g\perp}(\eta)$
from the following consideration. At $\kappa < 0$ one has 
$R_{g\perp} \sim a N^{\nu}$, so that ${\cal G}_{g\perp} = {\rm const}$.
In the opposite limit $\eta \gg 0$ the $N$-dependence drops out
and ${\cal G}_{g\perp} (\eta) \sim \eta ^{-\nu/\phi}$.
Thus
\begin{eqnarray}
{\cal G}_{g\perp} (\eta) =
\begin{cases} 
{\rm const} &\quad, \quad {\rm for} \quad \eta \le 0\\
\eta^{-\nu/\phi} &\quad, \quad {\rm for} \quad \eta \gg 0
\end{cases}
\label{Scaling_func_perp}
\end{eqnarray}
As a result 
\begin{eqnarray}
R_{g\perp} (\eta) \propto 
\begin{cases} 
a N^{\nu} &\quad, \quad {\rm for} \quad \eta \le 0\\
\kappa^{- \nu/\phi} &\quad, \quad {\rm for} \quad \eta \gg 0
\end{cases}
\label{Perp}
\end{eqnarray}

The gyration radius in direction parallel to the surface has 
similar scaling representation:
\begin{eqnarray}
R_{g\parallel} (\eta) = a N^{\nu} {\cal G}_{g\parallel} \left(
\eta\right)
\label{Rad_parall}
\end{eqnarray}
Again at $\kappa < 0$ the gyration radius $R_{g\parallel} \sim a N^{\nu}$ and
${\cal G}_{g\parallel} = {\rm const}$. At $\eta \gg 0$ the chain behaves as a
two-dimensional self-avoiding walk (SAW), i.e. $R_{g\parallel} \sim a
N^{\nu_2}$, where $\nu_2=3/4$ denotes the Flory exponent in two dimensions. In
result, the scaling function behaves as
\begin{eqnarray}
{\cal G}_{g\parallel} (\eta) =
\begin{cases} 
{\rm const} & \quad, \quad {\rm at} \quad \eta \le 0\\
\eta^{(\nu_2 - \nu)/\phi} &\quad, \quad {\rm at} \quad \eta \gg 0
\end{cases}
\label{Scaling_func_parall}
\end{eqnarray}
Thus
\begin{eqnarray}
R_{g\parallel} (\eta) \propto \begin{cases} a N^{\nu} &\quad, \quad {\rm at}
\quad \eta \le 0\\
\kappa^{(\nu_2 - \nu)/\phi} N^{\nu_2}&\quad, \quad {\rm at} \quad \eta \gg 0
                   \end{cases}
\label{Parall}
\end{eqnarray}

\subsubsection{Blob picture}

In the limit $\kappa N^{\phi} \gg 1$ the adsorbed chain can be visualized 
as a string of {\it adsorption blobs} which forms a pancake-like 
quasi-two-dimensional layer on the surface. The blobs are defined to contain 
as many monomers $g$ as necessary to be on the verge of being 
adsorbed and therefore carry an adsorption energy of the order of $k_B T$ 
each. The thickness of the pancake $R_{g\perp}$ corresponds to the size of
the blob and the chain conformation within a blob stays unperturbed (i.e.
it is simply a SAW), thus
$g \sim \left( R_{g\perp}/a\right)^{1/\nu} = \kappa^{-1/\phi} $ where we
have used eq~\ref{Perp}. The gyration radius can be represented thus as
\begin{eqnarray}
R_{g\parallel}  = R_{g\perp} \left( \frac{N}{g}\right)^{\nu_2} 
\propto \kappa^{(\nu_2 - \nu)/\phi} N^{\nu_2}
\label{Blob}
\end{eqnarray}
and one goes back to eq~\ref{Parall} which proves the consistency of the
adsorption blob picture. Generally speaking, the number of blobs,
$N/g \sim \kappa^{1/\phi} N$, is essential for the main scaling argument  
in the above-mentioned scaling functions. For example we could recast 
the order parameter scaling behavior eq~\ref{Scaling_first} as
\begin{eqnarray}
f = N^{\phi - 1} H \left( \frac{N}{g}\right) 
\label{Scaling_second}
\end{eqnarray}
where $H (x)$ denotes a new scaling function :
\begin{eqnarray}
H (x) = 
\begin{cases} 
{\rm const} &\quad, \quad {\rm for} \quad x \rightarrow 0\\
x^{1 - \phi} &\quad, \quad {\rm for} \quad x \gg 1
\end{cases}
\label{In_terms_blob}
\end{eqnarray}

\subsubsection{Ratio of gyration radius components}

The study of the ratio, $r (\eta) \equiv R_{g\perp}/R_{g\parallel}$, of 
gyration radius components is a convenient way to find the value of $\epsilon_c$
(see \cite{Sommer,Metzger}). In fact, from the previous scaling equations
\begin{eqnarray}
r (\eta) \equiv \frac{R_{g\perp}(\eta)}{R_{g\parallel}(\eta)} 
= \frac{{\cal G}_{g\perp}(\eta)}{{\cal G}_{g\parallel}(\eta)}
\label{Ratio}
\end{eqnarray}
Hence at the critical point, i.e. at $\eta \rightarrow 0$, the ratio $r (0) =
const$ is independent of $N$. Thus by plotting $r$ vs. $\epsilon$ for different
$N$ all such curves should intersect at a single point which gives $\epsilon_c$.

Another way to fix $\epsilon_c$ is the following. Exactly at the critical point
$f \sim N^{\phi - 1}$, so that by plotting $f N^{1 - \phi}$ vs. $N$ at different
values of $\epsilon$ one can determine the value $\epsilon \approx  \epsilon_c$
under which $f N^{1 - \phi}$ becomes independent of $N$.

\subsection{Free energy of adsorption}

The adsorption on a surface at $\kappa > 0$ is due to a free energy gain which
is proportional to the number of blobs, i.e.,
\begin{eqnarray}
\frac{F - F_{\rm bulk}}{N} \propto - \frac{1}{g} \sim - \kappa^{1/\phi} \;.
\label{Free_energy}
\end{eqnarray}
The expression for the specific heat per monomer follows immediately from
eq~\ref{Free_energy} as
\begin{equation}\label{cv}
C_V =- \frac{\partial^2 (F - F_{\rm bulk} )}{\partial^2 \kappa }
\propto \kappa^{-\alpha}
\end{equation}
where $\alpha = 2 - \phi^{-1}$.
Note that a factor of $k_BT$ is absorbed in the free
energy throughout the
paper. If $\phi = 0.5$ then $\alpha = 0$ and the specific heat undergoes a jump
at the CAP (cf. Section~\ref{OP_variance}). 

For a  chain (of the length $N$) on the verge of adsorption, the foregoing
free energy gain, $F - F_{\rm bulk}$, should be of the order of unity. In view
of eq~\ref{Free_energy} this gives an estimate for the critical energy of
adsorption - CAP,
\begin{eqnarray}
\epsilon_{c}(N) = \epsilon_{c}(\infty) \left( 1 + \frac{1}{N^{\phi}}\right),
\label{Fin_epsilon}
\end{eqnarray}
where we have explicitly marked the CAP, $\epsilon_{c}(N)$ and
$\epsilon_{c}(\infty)$, for finite and infinitely long chains respectively.

 \section{Multi-block copolymer adsorption}

Consider now the adsorption of a regular multi-block copolymer which is built up
from monomers $A$ which attract (stick) to the substrate and monomers $B$
which are neutral to the substrate. In order to treat the adsorption of a
regular multi-block $AB$ - copolymer we reduce the problem to that of a
homopolymer  which has been considered above. The idea is that a regular
multi-block copolymer can be considered as a ``homopolymer'' where a single
$AB$-diblock plays the role of an effective monomer~\cite{Corsi}. For such a
mapping we first estimate the effective energy of adsorption per diblock.

\subsection{Effective energy of adsorption per diblock}

\begin{figure}[ht]
\begin{center}
\includegraphics[scale=0.5]{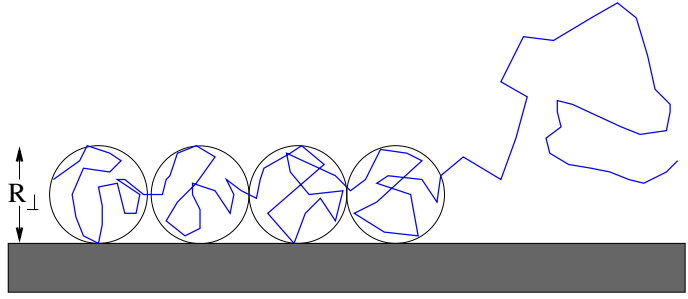}
\caption{Schematic representation of an individual adsorbed $AB$-diblock. The
$A$ - part forms a string of quasi-two dimensional blobs and the $B$-part is
neutral regarding the substrate and its contribution to the free energy is of
pure entropical nature.}
\label{Tail}
\end{center}
\end{figure}

Each individual diblock is made up of an attractive $A$-block of length $M$
and a neutral $B$-block of the same length $M$. Upon adsorption the
attractive $A$-block forms a string of blobs whereas the $B$-part forms a
non-adsorbed tail (or loop) - (see Figure~\ref{Tail}).
The free energy gain of the attractive block may be written according to
eq~\ref{Free_energy} as
\begin{eqnarray}
F_{\rm attr} = - \kappa^{1/\phi} M
\end{eqnarray}
where we measure the energy in units of $k_B T$ and
$\kappa \equiv (\epsilon - \epsilon^h_c)/\epsilon^h_c$ measures the normalized
distance from the CAP $\epsilon^h_c$ of a homopolymer. The neutral $B$-part
which is most frequently a loop connecting adjacent $A$-blocks, but could also
be a tail with the one end free, contributes only to the entropy loss
\begin{eqnarray}\label{gamma}
F_{\rm rep} = (\gamma - \gamma_{11}) \ln M
\end{eqnarray}
where the universal exponents $\gamma$ and $\gamma_{11}$ are well
known\cite{Vander} (e.g. in 3$D$ - space $\gamma = 1.159 $, $\gamma_{11} =
-0.390$). In case that also the tails are involved, one should also use the
exponent $\gamma_1 = 0.679$ albeit this does not change qualitatively the
expression eq~\ref{gamma}. They enter the partition function expressions for a
free chain, a chain with both ends fixed at a two points, and for a chain,
tethered by the one end~\cite{Vander}. In result the effective adsorption
energy of a diblock
is
\begin{eqnarray}
E (M) = \kappa^{1/\phi} M  - (\gamma - \gamma_{11}) \ln M
\label{Attraction}
\end{eqnarray}

\subsection{Order parameter}

Now we consider a 'homopolymer' which is build up from effective units
(diblocks), with the attractive energy given by eq~\ref{Attraction}. Let us
denote the total number of such effective units by ${\cal N} = N/2M$. The
fraction of effective units on the surface obeys then the same scaling law
as given by eq~\ref{Scaling_first}, i.e.,
\begin{eqnarray}
\frac{{\cal N}_s}{{\cal N}} =
{\cal N}^{\phi - 1} G \left( \Delta {\cal N}^{\phi}\right) 
\label{Fraction_phi}
\end{eqnarray}
where now $\Delta \equiv (E - E^h_c)/E^h_c$ with the critical adsoption energy
$E^h_c$ of the renormalized homopolymer. Generally, one would expect $E^h_c$
to be of the order of $\epsilon^h_c$ albeit for different models both
critical energies would probably differ from each other. 
Eq~\ref{Fraction_phi} is accurate if one require that (i) $\kappa \ll 1$
but $M \gg 1$ such that $\ln M \gg 1$ and $\kappa^{1/\phi}M \gg 1$, 
and (ii) ${\cal N } \gg 1$. The effective
attraction $E$ of a segment of the renormailzed chain now depends on $M$
according to eq~\ref{Attraction}.
 
Within each effective unit only $M_s$ $A$-monomers will be adsorbed at
criticality whereby this monomer number scales as
\begin{eqnarray}
M_s = M^{\phi} G \left( \kappa M^{\phi}\right) 
\label{M_s}
\end{eqnarray}
with $\kappa \equiv (\epsilon - \epsilon^h_c)/\epsilon^h_c$. 
 
The total number of adsorbed monomer is given by 
\begin{eqnarray}
N_s = {\cal N}_s  M_s = {\cal N}_s  M^{\phi} G \left( \kappa M^{\phi}\right)
\end{eqnarray}
It follows that the fraction
\begin{eqnarray}
f \equiv \frac{N_s}{N} &=& \frac{{\cal N}_s}{N} \: M^{\phi} G \left( \kappa
M^{\phi}\right) = \frac{{\cal N}_s}{2 {\cal N}} \: M^{\phi - 1} G \left( \kappa
M^{\phi}\right)\nonumber\\
&=& \frac{1}{2}\: M^{\phi - 1} G \left( \kappa M^{\phi}\right) \: \left(
\frac{N}{2 M}\right)^{\phi - 1} \: G \left( \Delta \left(
\frac{N}{M}\right)^{\phi}\right) \quad,
\end{eqnarray}
where we have used the scaling law, eq~\ref{Fraction_phi}, for the effective
units. Hence, the final expression for the order parameter can be written as
follows:
\begin{eqnarray}
f = \frac{1}{2^{\phi}}\: N^{\phi - 1} \: G \left( \kappa M^{\phi}\right) 
\:  G \left( \Delta \left( \frac{N}{M}\right)^{\phi}\right) 
\label{Order_parameter}
\end{eqnarray}
Thus we have expressed the order parameter $f$ of a multi-block
copolymer in terms of the chain length $N$, the block length $M$, the monomer 
attraction energy $\epsilon$ as well as the model-dependent homopolymer
critical attraction energy $\epsilon^h_c$. Let us consider now some limiting
cases.

\subsubsection{Close to criticality $\Delta = 0$}

At the CAP of the multiblock chain one has $\Delta = 0$, thus one can estimate
the deviation $\kappa^M_c$, of the corresponding critical energy of adsorption,
$\epsilon^M_c$, from that of a homopolymer, namely
 \begin{eqnarray}
\kappa^M_c \equiv \frac{\epsilon^M_c - \epsilon^h_c}{\epsilon^h_c}  = \left(
\frac{(\gamma - \gamma_{11}) \ln M + E^h_c}{M}\right)^{1/2}
\label{Kappa_vs_M}
 \end{eqnarray}
where we have used eq~\ref{Attraction} and set $\phi = 0.5$.  Under this
condition the second $G$ -function in eq~\ref{Order_parameter} is a constant,
i.e., $G (0) = {\rm const}$. On the other hand, with respect to a single
effective unit the chain stays far from the criticality because of $\kappa^M_c
\sqrt{M} =\sqrt{(\gamma - \gamma_{11}) \ln M + E^h_c}\gg 1$. In this case
the first $G$ - function in eq~\ref{Order_parameter} behaves as $G (\kappa^M_c
\sqrt{M}) \sim \kappa^M_c \sqrt{M}$ where $\kappa^M_c$ now is fixed by 
eq~\ref{Kappa_vs_M}. In result, eq~\ref{Order_parameter} becomes
\begin{eqnarray}
f \propto \left( \frac{(\gamma - \gamma_{11}) \ln M +
E^h_c}{N}\right)^{1/2}
\label{Order_parameter_critical}
\end{eqnarray}

\subsubsection{State of the strong adsorption}

In this regime $\kappa \sqrt{M} \gg 1$ and $\Delta \sqrt{N/M} \gg 1$ so that
$f \simeq (1/\sqrt{N}) G(\kappa \sqrt{M})G(\Delta \sqrt{N/M}) \sim \kappa
\Delta$. Therefore, 
\begin{eqnarray}
f \simeq \frac{\kappa \left[\kappa^2 M - (\gamma - \gamma_{11}) 
\ln M - E^h_c \right] }{E^h_c}
\label{Order_parameter_deep}
\end{eqnarray}

\subsection{Gyration radius}

The components of the gyration radius of a multi-block copolymer can be
treated again by making use of the mapping on the homopolymer problem given by
eqs~\ref{Rad_perp} and \ref{Rad_parall}. In doing so the mapping looks 
as follows:
\begin{eqnarray}
a &\longrightarrow& a M^{\nu}\nonumber\\
\kappa &\longrightarrow& \Delta = \frac{E - E^h_c}{E^h_c}\\
N &\longrightarrow& {\cal N} = \frac{N}{2 M}\nonumber
\label{Map}
\end{eqnarray}
Thus the gyration radius component in direction perpendicular to the surface
becomes
\begin{eqnarray}
{\cal R}_{g\perp} = a N^{\nu} \: {\cal G}_{g\perp} 
\left( \Delta \left( \frac{N}{M}\right)^{\phi}\right) 
\label{Rad_copol_perp}
\end{eqnarray}
In the strong adsorption limit $\Delta \sqrt{N/M} \gg 1$  and ${\cal R}_{\perp}
\sim a \Delta^{- \nu/\phi} M^{\nu}$, which yields
\begin{eqnarray}
{\cal R}_{\perp} \simeq   \frac{a M^{\nu}  {E^h_c}^{2 \nu}}{\left[
\kappa^2 M -(\gamma - \gamma_{11}) \ln M - E^h_c\right]^{2 \nu} }
\label{Copoly_perp}
\end{eqnarray}

In a similar manner, the gyration radius component parallel to the surface has
the form
\begin{eqnarray}
{\cal R}_{g\parallel} = a N^{\nu} \: {\cal G}_{g\parallel} \left( \Delta \left(
\frac{N}{M}\right)^{\phi}\right)
\label{Rad_copol_parallel}
\end{eqnarray}
which in the limit $\Delta \sqrt{N/M} \gg 1$ results in 
\begin{eqnarray}
{\cal R}_{g\parallel} &\simeq& a \left( \frac{\Delta^{1/\phi}}{M}
\right)^{\nu_2 - \nu} \: N^{\nu_2}\nonumber\\
&\simeq& \frac{a \left[ \kappa^2 M - (\gamma - \gamma_{11}) \ln M 
- E^h_c\right]^{2(\nu_2 - \nu)} }{M^{\nu_2 - \nu}}
\: N^{\nu_2}
\end{eqnarray}

Like in the homopolymer case, one can define a blob length
$g_{\rm eff} \sim \left({\cal R_{\perp}}/a\right)^{1/\nu} \sim \Delta^{-1/\phi} 
\: M$ which in the strong adsorption limit, $\Delta \geq 1$, approaches
the block length, $g_{\rm eff} \simeq M$, as it should be.

Also in the limit of strong adsorption, $\Delta \sqrt{N/M} \gg 1$, the ratio
\begin{eqnarray}
\frac{{\cal R}_{g\parallel}}{{\cal R}_{\perp}} \simeq \left( 
\frac{\Delta^{1/\phi} N}{M}\right)^{\nu_2} \simeq 
\left( \frac{N}{g_{\rm eff}}\right)^{\nu_2}
\end{eqnarray}
leads to the correct scaling in terms of number of blobs.

\section{Random copolymer adsorption}

Consider a random copolymer which is built up of $N_p$ $A$-type and $N_h$
$B$-type monomers. The sampled $AB$-sequences are frozen (i.e. a distinct sample
does not change during the measurement) which corresponds to quenched disorder.
The binary variable $\sigma$ specifies the arrangement of monomers along the
chain, so that $\sigma = 1$, if the monomer is of $A$-type ($A$-monomers attract
to the surface) and $\sigma =0$ otherwise (i.e. in case of neutral
$B$-monomers). Let the fraction of attractive monomers (i.e., the composition)
be $p = N_p/N$ and the fraction of neutral ones be $1 - p = N_h/N$. We assume
that the statistics of sequences is governed by the Bernoulli
distribution~\cite{Odian}, i.e., the corresponding distribution function looks
like:
\begin{eqnarray}
P\left\lbrace \sigma \right\rbrace  = p \delta (1 - \sigma) + (1 - p) \delta
(\sigma)
\label{Distribution}
\end{eqnarray}
This distribution is a special case of the more general Markovian
copolymers~\cite{Odian} when the "chemical correlation length" goes to zero. Two
statistical moments which correspond to the distribution eq~\ref{Distribution}
are
\begin{eqnarray}
\left\langle \sigma \right\rangle  &=& p\nonumber\\
\left\langle \theta^2 \right\rangle  &\equiv& \left\langle 
\left[ \sigma - \left\langle \sigma \right\rangle \right]^2
\right\rangle  = p (1 - p)
\end{eqnarray}

\subsection{How does the critical $\epsilon_c$ depend on the
composition $p$?}

The adsorption of a random copolymer on a homogeneous surface has been studied
by Whittington et al.~\cite{Whit_1,Whit_2} within the framework of the annealed
disorder approximation. Physically this means that during the measurements the
chain touches the substrate at random in such a way that, as a matter of fact,
one samples all possible distributions of monomers sequences along the backbone
of the macromolecule. Following this assumption~\cite{Whit_1}, let $c_{N}^{+}
(n)$ be the number of  polymer configurations such that $n$ units have contact
with the surface simultaneously. The percentage of $A$-monomers (composition) is
denoted by $p$. In the annealed approximation one then averages the partition
function over the disorder distribution, i.e.,
\begin{eqnarray}
Z (\epsilon) &=& \sum\limits_{n=1}^{N} \sum\limits_{n_p = 0}^{n}
\: c_{N}^{+} (n) \: \left( {n \atop n_p}\right)  p^{n_p} (1-p)^{n-n_p}
\: {\rm e}^{\epsilon n_p}\nonumber\\
&=&\sum\limits_{n=1}^{N} c_{N}^{+} (n) \: \left[ p {\rm e}^{\epsilon}+1
- p\right]^{n} = \sum\limits_{n=1}^{N} c_{N}^{+} (n) 
\: {\rm e}^{n \:\epsilon^{h}_{\rm eff}}
\label{Annealed}
\end{eqnarray}
where $\epsilon^{h}_{\rm eff}$ is the attraction energy of an effective
homopolymer. From eq~\ref{Annealed} one can see that the annealed problem is
reduced to that of a homopolymer where the effective attractive energy is
defined as
\begin{eqnarray}
\epsilon^{h}_{\rm eff} = \ln \left[p {\rm e}^{\epsilon} +1 - p\right]
\label{Homo}
\end{eqnarray}
We know that at the critical point the homopolymer attraction energy, 
$\epsilon^{h}_{\rm eff} = \epsilon^{h}_{c}$, is model dependent.
Then the critical attraction energy $\epsilon=\epsilon_{c}^p$ of a
random copolymer reads
\begin{eqnarray}
\epsilon_{c}^p = \ln \left[ \frac{\exp{\epsilon^{h}_{c} + p - 1}}{p}
\right] \geq \epsilon^{h}_{c}
\label{Copol}
\end{eqnarray}
where the composition $0 \leq p \leq 1$. At $p \rightarrow 0$  $\epsilon_{c}^p
\rightarrow \infty$ whereas at $p=1$ $\epsilon_c^p = \epsilon^{h}_{c}$.
The relationship in eq~\ref{Copol} has been recently found to be confirmed by
Monte Carlo simulations~\cite{Ziebarth}.

\section{Simulation Methods}

\begin{figure}[ht]
\begin{center}
\includegraphics[scale=0.62]{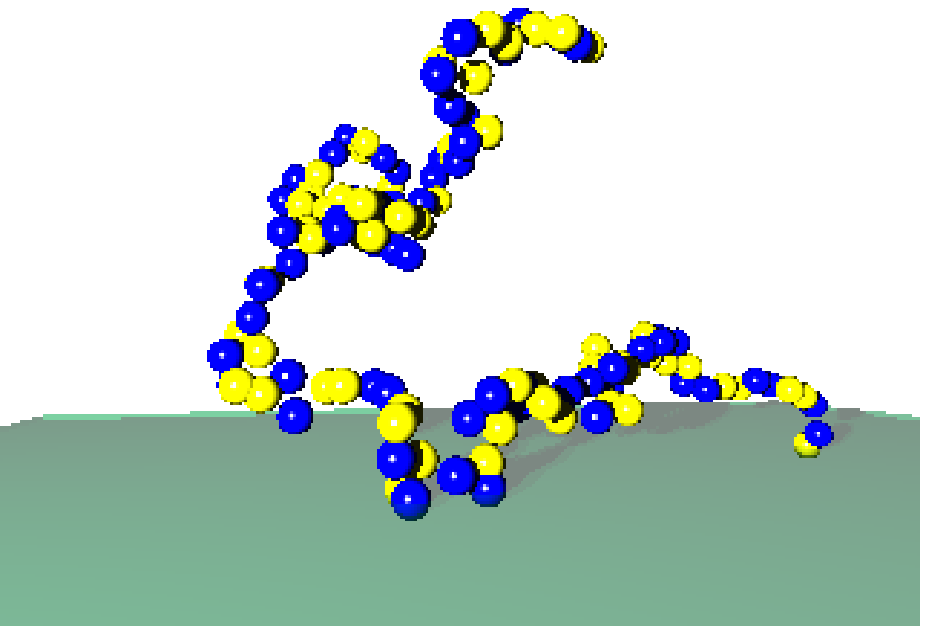}\hspace{1.0cm}
\includegraphics[scale=0.42, angle=0]{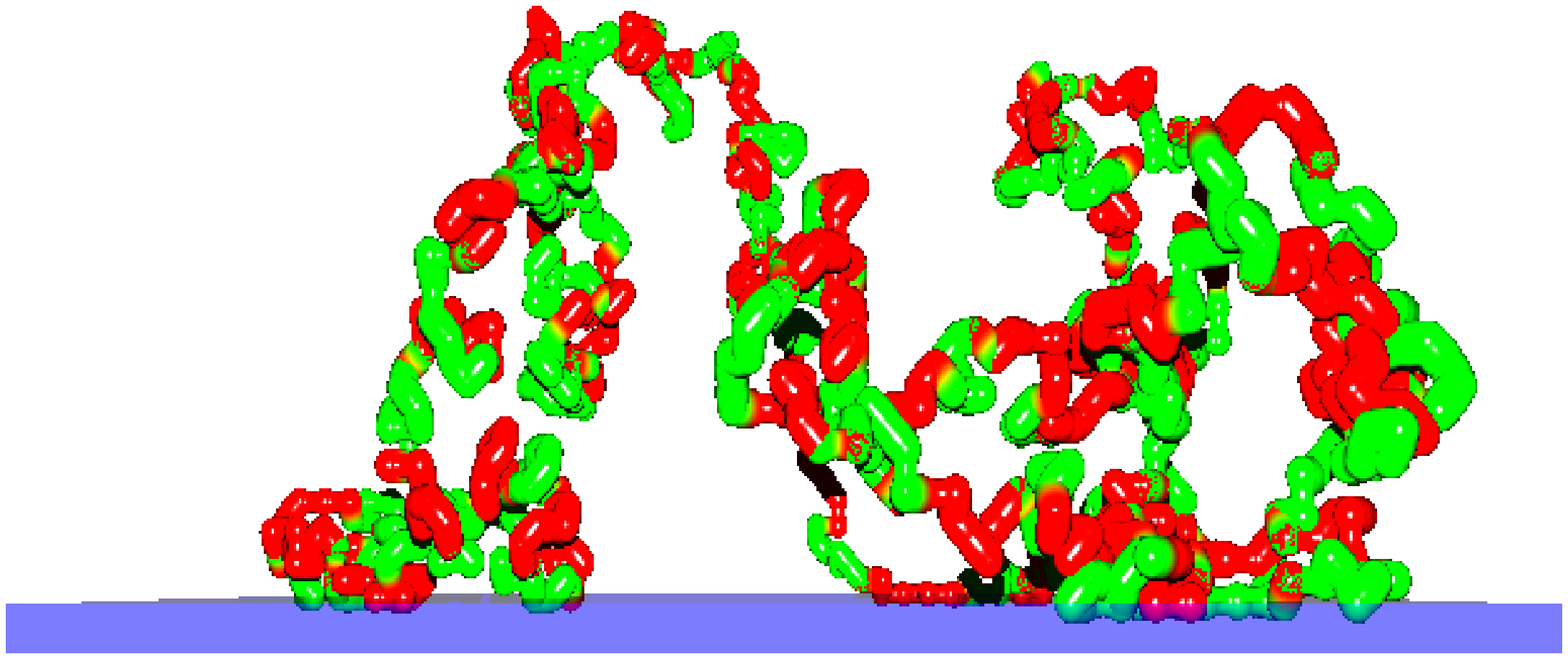}
\caption{Schematic representation of a grafted chain close to criticality.
(a) Snapshot of a chain with length $N=128$ from the MA model and block size
$M=2$; (b) $N=2048$ with $M=8$ from the PERM simulation.}
\label{Criticality}
\end{center}
\end{figure}

To check the theoretical predictions mentioned in the previous sections
we have performed Monte Carlo simulations and investigated the adsorption of a
homopolymer, multi-block copolymers, and random copolymers on flat surfaces. 
Two coarse-grained models, the bead spring model and the simple cubic lattice
model, Figure~\ref{Criticality}, are used, and two different Monte Carlo
algorithms, the Metropolis algorithm (MA) and pruned-enriched Rosenbluth method
(PERM), are applied to the two models, respectively.

\subsection{Off-lattice bead spring model with MA}

We have used a coarse grained off-lattice bead spring model\cite{AMKB} to
describe the polymer chains. Our system consists of a single chain tethered at
one end to a flat structureless surface. There are two kinds of monomers: "A"
and "B", of which only the "A" type feels an attraction to the surface. The
surface interaction of the "A" type monomers is described by a square well
potential $U_w(z) =\epsilon $ for $z<\delta$ and $U_w(z) =0$ otherwise. Here
$\epsilon /k_BT$ is varied from $0.6$ to $3.6$. The effective bonded interaction
is described by the FENE (finitely extensible nonlinear elastic) potential.
\begin{equation}
U_{FENE}= -K(1-l_0)^2ln\left[1-\left(\frac{l-l_0}{l_{max}-l_0} \right)^2 \right]
\label{fene}
\end{equation}
with $K=20, l_{max}=1, l_0 =0.7, l_{min} =0.4$

The nonbonded interactions are described by the Morse potential.
\begin{equation}
\frac{U_M(r)}{\epsilon_M} =\exp(-2\alpha(r-r_{min}))-2\exp(-\alpha(r-r_{min}))
\end{equation}
with $\alpha =24,\; r_{min}=0.8,\; \epsilon_M/k_BT=1$.

We use periodic boundary conditions in the $x-y$ directions and impenetrable
walls in the $z$ direction. We have studied polymer chains of lengths $32$,
$64$, $128$, $256$ and $512$. We have also studied homopolymer chains and
random copolymers (with a fraction of attractive monomers,
$p=0.25,\;0.5,\;0.75$). The size of the box was $64\times 64\times 64$ in all
cases except for the $512$ chains where we used a larger box size of $128\times
128\times 128$. The standard Metropolis algorithm was employed to govern the
moves with  self avoidance automatically incorporated in the potentials. In each
Monte Carlo update, a monomer was chosen at random and a random displacement
attempted with $\Delta x,\;\Delta y,\;\Delta z$ chosen uniformly from the
interval $-0.5\le \Delta x,\Delta y,\Delta z\le 0.5$. The transition probability
for the attempted move was calculated from the change $\Delta U$ of the
potential energies before and after the move as $W=exp(-\Delta U/k_BT)$. As for
standard Metropolis algorithm, the attempted move was accepted if $W$ exceeds a
random number uniformly distributed in the interval $[0,1]$.

\subsection{Coarse-grained lattice model with PERM}

The adsorption of $AB$ block copolymer with one end (monomer $A$) grafted to a
plane impenetrable surface and with only monomers $A$ attractive to the surface
are described by SAWs of $N-1$ steps on a simple cubic lattice with restriction
$z \ge 0$. There is an attractive interaction between monomers $A$ and the wall.
The partition sum now is written as
\begin{equation} 
     Z_N^{(1)}(q) =\sum_{N_s} A_N(N_s) q^{N_s}   \label{ZNA}
\end{equation}
where $A_N(N_s)$ is the number of configurations of SAWs with $N$ steps having
$N_s$ sites on the wall, and $q=e^{\epsilon/k_BT}$ ($k_BT=1$ hereafter) is the
Boltzmann factor, $\epsilon>0$ is the attractive energy between the monomer $A$
and the wall. As $q \rightarrow 1$, there is no attraction between the monomer
$A$ and the wall. On the other hand it becomes clear that any copolymer will
collapse onto the wall, if $q$ becomes sufficiently large. Therefore we expect a
phase transition from a grafted but otherwise detached to an adsorbed phase,
similar to the transition observed also for homopolymers.

   For our simulations, we use the pruned-enriched Rosenbluth method
(PERM)~\cite{G97} which is a biased chain growth algorithm with resampling
("population control") and depth-first implementation. Polymer chains are built
like random walks by adding one monomer at each step. Thus the total weight of a
configuration for a polymer consisting of $N$ monomers is a product of those
weight gains at each step, i.e. $W_N=\Pi_{i=0}^{N-1} w_i$. As in any such
algorithm, there is a wide range of possible distributions of sampling, we
have the freedom to give a bias at each step while the chain grows, and the bias
is corrected by means of giving a weight to each sample configuration, namely,
$w_i \rightarrow w_i/p_i$ where $p_i$ is the probability for putting the monomer
at step $i$. In order to suppress the fluctuations of weights as the chain is
growing, the population control is done by "pruning" configurations with too low
weight and "enriching" the sample with copies of high-weight configurations.
Therefore, two thresholds are introduced here, $W_n^+=c^+ Z_n$ and $W_n^-=c^-
Z_n$, where $Z_n=\frac{1}{M_n}\sum_{config.} W_n$ from the $M_n$ trail
configuration is the current estimate of partition sum at the $n-1$ step, $c^+$
and $c^-$ are constants of order unity and $c^+/c^- \approx 10$.
   In order to compare with the results obtained by the first method, we
simulate homopolymers of length $N=2048$ and multi-block copolymers with
block size $M=2^k\;,$ $k=0,1,2,\cdots,9$. The number of monomers is increased
to $N=8192$ as the block size increases. Also random copolymers of $N=2048$
monomers with composition $p=0.125$, $0.25$, $0.50$, and $0.75$ are sampled. 

\section{Simulation Results}

\subsection{Determination of the critical point of adsorption}

The determination of the critical adsorption point (CAP) is essential for
testing the scaling results and for comparison with theory. In this work we
determine the CAP from the analysis of several quantities: the order parameter
$f$, the variance of the number of adsorbed monomers, $C_v$, and the gyration
radius $R_g$. These methods are described as follows: \\
\subsubsection{CAP from the order parameter}\label{OP_CAP}
\begin{figure}[bht]
\begin{center}
\vspace{1cm}
\includegraphics[scale=0.32]{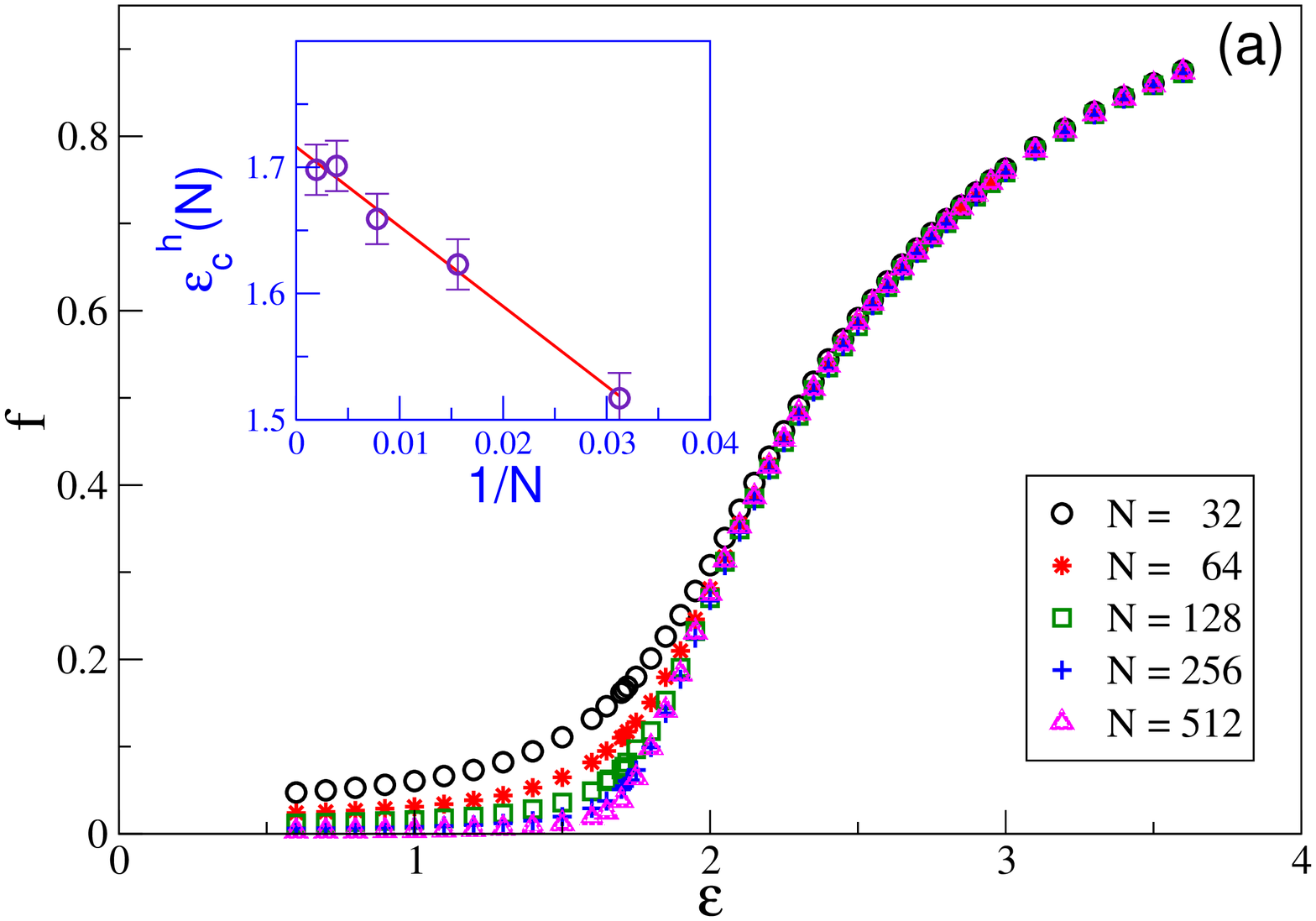}\hspace{0.7cm}
\includegraphics[scale=0.32]{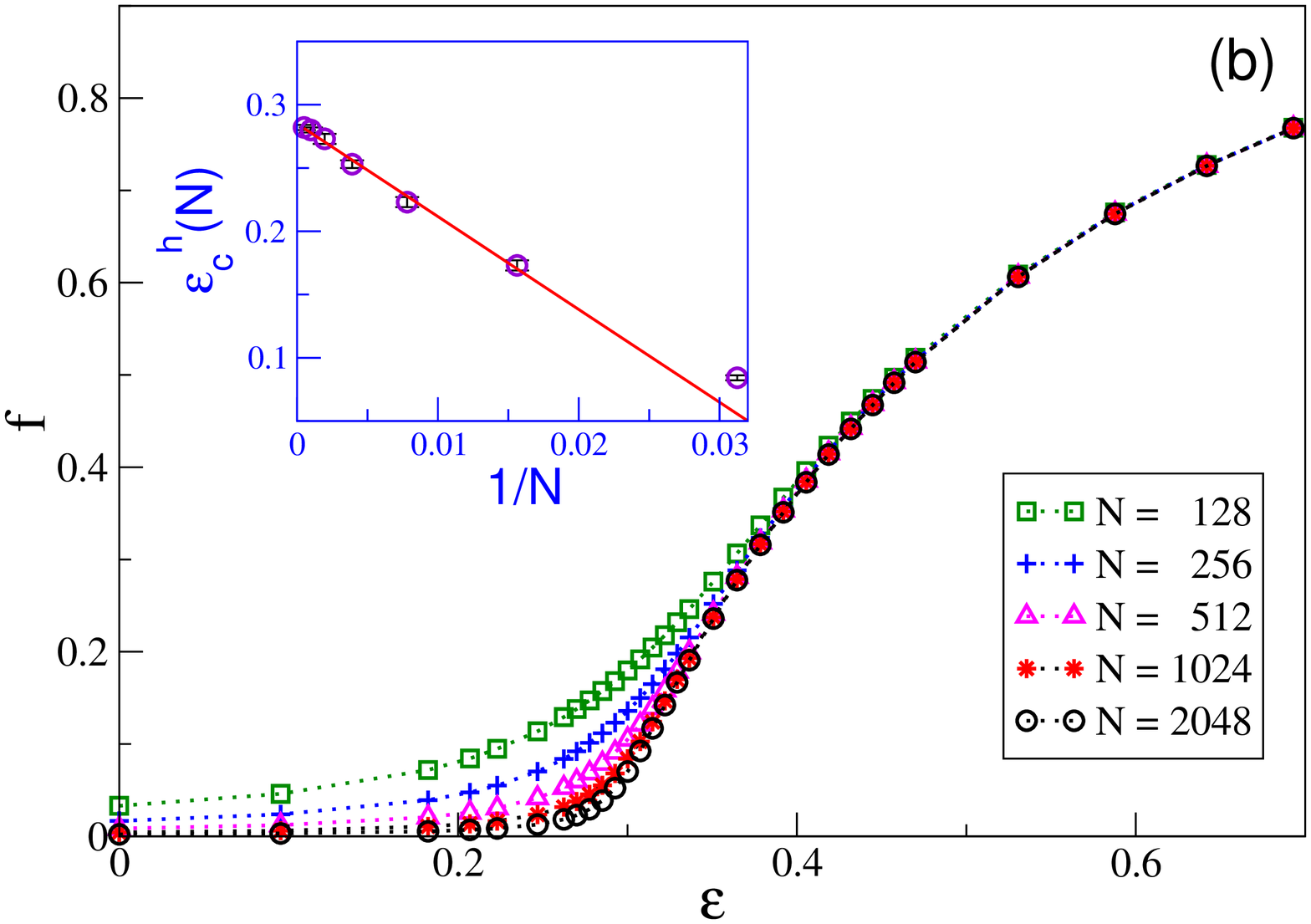}
\caption{The order parameter $f$ against the adsorption energy $\epsilon$ for
homopolymers of different chain lengths $N$. The value of the CAP
$\epsilon_c^h(N)$ for $N\rightarrow \infty$ is extrapolated from the log-log
plot of $\epsilon_c^h(N)$ versus $1/N$ as shown in the insert. In the
thermodynamic limit (a) $\epsilon^h_c \approx 1.716$ (MA off-lattice model), (b)
$\epsilon^h_c \approx 0.284$ (PERM on a cubic lattice).}
\label{f_homo}
\end{center}
\end{figure}

\begin{figure}[bht]
\vspace{1cm}
\includegraphics[scale=0.32]{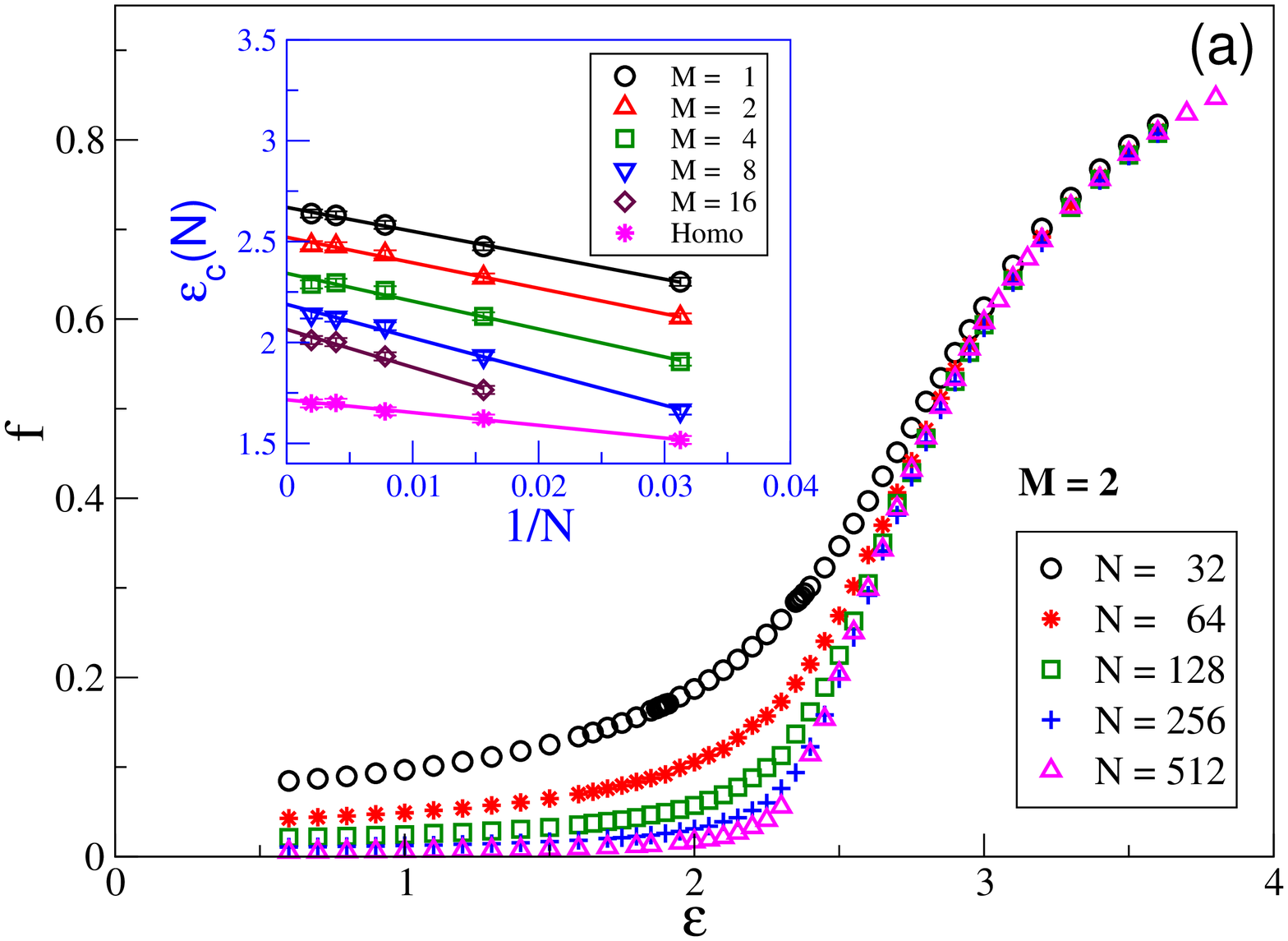}\hspace{0.7cm}
\includegraphics[scale=0.32]{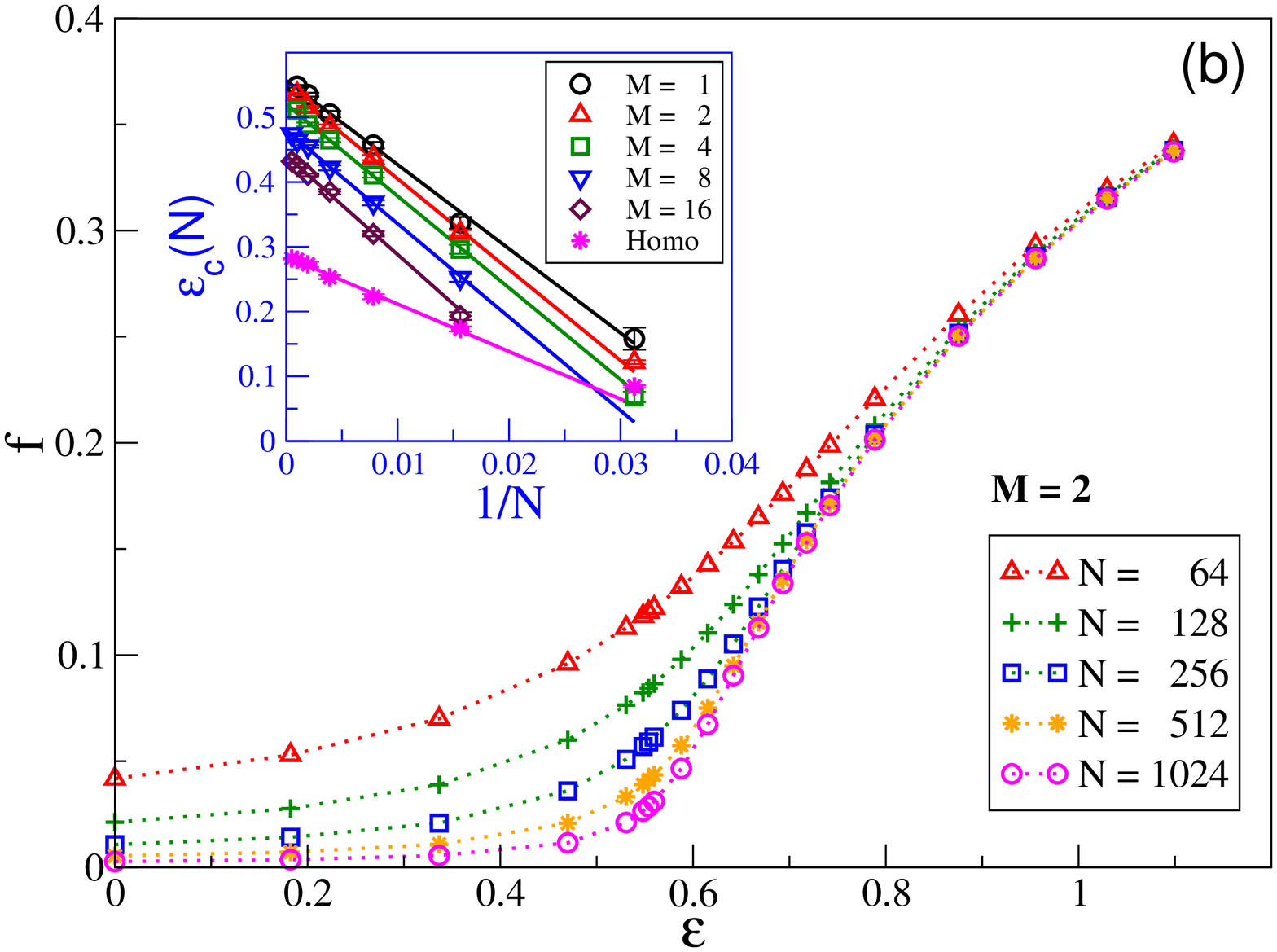}
\caption{The order parameter $f$ plotted as a function of attractive energy
$\epsilon$ for copolymers with block size $M=2$. The extrapolation plots for
$\epsilon_c(N)$ versus $1/M$ for block sizes $M=1$, $2$, $4$, $8$, $16$, and for
the homopolymer, plotted versus $1/N$, are shown
in the insert. (a) - the MA model, (b) - PERM.}
\label{f_m2}
\end{figure}

From the plots of the order parameter $f$
against the adsorption energy $\epsilon$ for chains of different length $N$ we
determine the CAP as the point where the tangent taken at the inflection point
of the order parameter curve intersects the horizontal axis $\epsilon$. Results
are shown in Figure~\ref{f_homo} for homopolymers and in Figure~\ref{f_m2} for
multi-block copolymer with block size $M=2$.
In Figure~\ref{f_homo}a and~\ref{f_m2}a data is obtained by MA method in our
off-lattice model, while in Figures~\ref{f_homo}b and~\ref{f_m2}b the data is
obtained by PERM for self-avoiding chains on a cubic lattice. Evidently, in both
cases the order parameter $f$ increases with growing strength of
the substrate potential $\epsilon$. Thus the polymer chain undergoes a
transition from a grafted, but otherwise detached state, to an adsorbed state
whereby the chain lies flat on the surface plane - see Figure~\ref{Criticality}b.
The transition region narrows down as $N$ increases, which is in good agreement
with the scaling prediction of $f$, eq~\ref{Fin_epsilon}, in all cases. In the
inset of Figures~\ref{f_homo} and Figure~\ref{f_m2}, we see that the critical
point $\epsilon_c^h(N)$ for homopolymers of chain length $N$ as well as the
critical points $\epsilon_c^M(N)$ for multi-block copolymers of chain length $N$
with $M=1$, $M=2$, $M=4$, $M=8$, and $M=16$, gradually increase as $N
\rightarrow \infty$. By extrapolating the data to $1/N=0$, one obtains the
CAP values in the thermodynamic limit. Results for $\epsilon^h_c$ by MA
and by PERM are listed in Tables~\ref{table1} and~\ref{table2}.
\begin{table}[htb]]{\caption{MA} \label{table1}
\begin{tabular}{|l|l|l|l|l|l|l|}
\hline
 M/N & 64 & 128 & 256 & 512  &  $\infty$, $f$ & $R_g$\\ \hline \hline
 1 & 2.47(3)  & 2.58(3)  & 2.63(3)   &2.63(3)   &2.672(30)   & 2.65(3)      \\
\hline
 2 & 2.32(3)  & 2.44(3)  &2.47(3)   &2.48(3)   &2.52(2)   &2.52(3)       \\
\hline
 4 & 2.13(3)  &2.260(3)   &2.29(3)   & 2.29(3)  &2.34(2)   & 2.30(4)     \\
\hline
 8 & 1.93(3)  &2.08(3)   &2.12(3)   &2.14(3)   & 2.19(3)  & 2.06(4)     \\
\hline
 16 &1.76(3)  & 1.93(3)  & 2.00(3)  &2.01(3)   &2.06(3)   & 1.95(4)     \\
\hline
p/N &  & & & &  & \\ \hline
1.0  &  1.62(2)  &1.66(2)   & 1.701(20)  &1.698(25)   &1.716(20)   &1.718(20)  
\\
\hline
0.75 &1.83(2)    &1.89(2)   &1.92(2)   & 1.946(20)  &1.95(3)   &1.95(3)   \\
\hline
0.50 & 2.21(2)   & 2.25(2)  & 2.29(2)  &2.32(2)   & 2.33(2)  & 2.38(5) \\ \hline
0.25 &  2.81(4)  & 2.97(4)  & 2.98(4)  &3.02(4)   & 3.05(5)   &2.91(6)   \\
\hline
\end{tabular}}
\end{table}
\begin{table}[htb]{\caption{PERM} \label{table2}
  \begin{tabular}{|l|l|l|l|l|l|l|l|}
\hline  
 M/N & 64 & 128 & 256 & 512 & 1024 &  $\infty$, $f$ & $R_g$\\ \hline \hline
 1 & 0.337(9) & 0.457(5) & 0.505(5) & 0.535(3) & 0.548(2) & 0.560(2) & 0.568(6) \\  \hline
 2 & 0.322(4) & 0.438(4) & 0.486(3) & 0.516(2) & 0.536(2) & 0.545(8) & 0.556(3) \\ \hline
 4 & 0.296(7) & 0.411(4) & 0.465(3) & 0.489(3) & 0.511(2) & 0.520(4) & 0.524(2) \\ \hline
 8 & 0.368(4) & 0.422(4) & 0.455(2) & 0.464(3) & 0.474(2) & 0.480(2) & 0.478(3) \\ \hline
 16 &0.320(4) & 0.385(4) & 0.411(2) & 0.426(3) & 0.432(2) & 0.441(2) & 0.437(4) \\ \hline
p/N &  & & & & & & \\ \hline
1.0  &  0.173(4) & 0.223(4) & 0.250(3) & 0.267(4) & 0.278(2) & 0.285(3) & 0.286(3) \\ \hline
0.75 &  0.241(10) & 0.294(6) & 0.325(5) & 0.346(3) & 0.352(3) & 0.363(2) & 0.366(2) \\ \hline
0.50 &  0.370(20) & 0.439(15) & 0.469(8) & 0.485(5) & 0.499(4) &  0.507(2) & 0.509(2) \\ \hline
0.25 &  0.77(2) & 0.78(2) & 0.82(1) & 0.83(2) &  0.83(2) & 0.843(6) & 0.845(4) \\ \hline
0.25/N &  100     &  200 & 400 & 800 & 1600 & & \\ \hline
 \end{tabular}
}
\end{table}
We should point out here that the simulation with the MA model requires
considerable computational effort for $N\ge 512$, therefore, with the PERM
method we confine ourselves to chain lengths not large than $N=2048$
(Figure~\ref{f_homo}b), which are considerably shorter than
feasible~\cite{Grassberger}. Nevertheless, our estimate of the CAP
$\epsilon_c^h=0.285(3)$ is in good agreement with previous
results~\cite{Grassberger} (within the error bars) although corrections to
scaling have not been considered here.

\subsubsection{From the variance of the order parameter:}
\label{OP_variance}
\begin{figure}[bht]
\begin{center}
\vspace{1cm}
\includegraphics[scale=0.32]{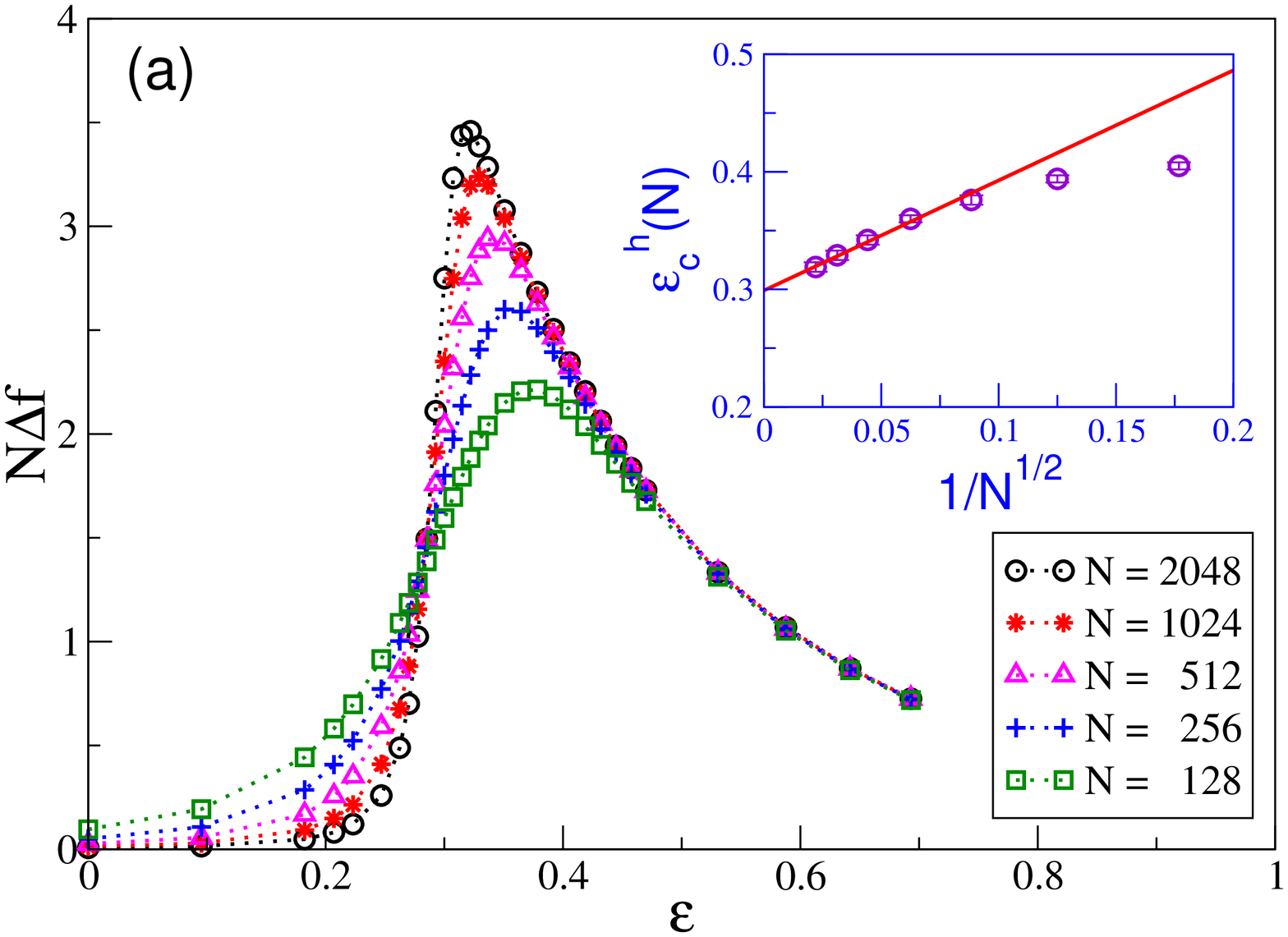}\hspace{0.7cm}
\includegraphics[scale=0.32]{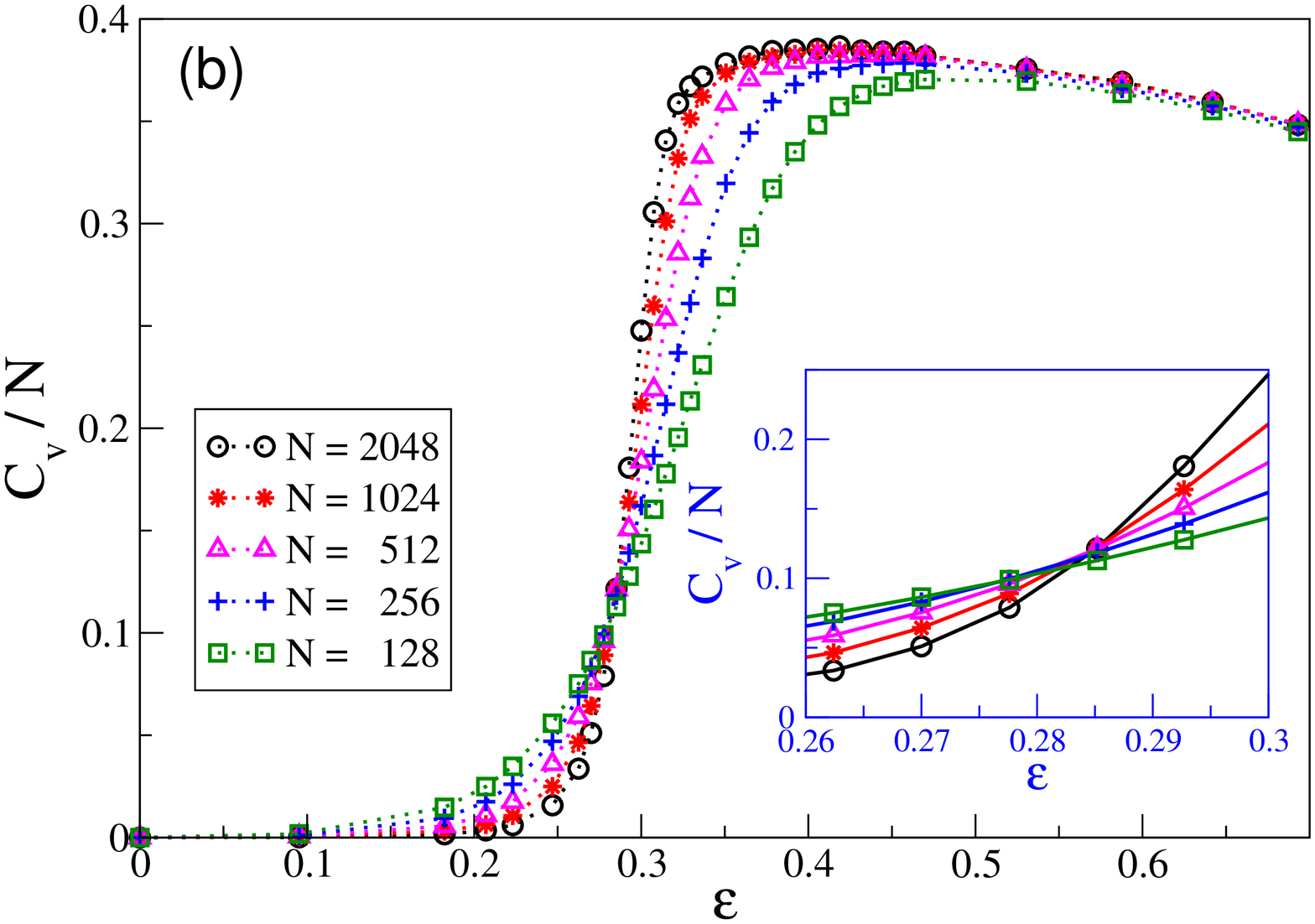}
\caption{Results for (a) the variance of the order parameter $f$ multiplied by
$N$ in the case of a homopolymer. The inset shows an extrapolation of the
CAP $\epsilon^h_c(N)$ for $1/N^{\phi} \rightarrow 0$ which converges to the
value for an infinite chain, $\epsilon^h_c = 0.284$ - cf. Table~\ref{table2}. 
(b) the specific heat per monomer, $C_V/N$, which plotted as a function of
$\epsilon$ for homopolymers of different chain length $N$.~\cite{Hsu}  }
\label{cv_homo_perm.eps}
\end{center}
\end{figure}

In a computer simulation one usually computes the variance of the order
parameter, $\Delta f$, which yields some important thermodynamic quantities like
isothermal compressibility, and/or specific heat, via the fluctuation relations.
\begin{eqnarray} \label{variance}
N^2 \Delta f = \langle N_s^2\rangle - \langle N_s\rangle ^2,
\end{eqnarray} 
At the CAP $\Delta f$ has a maximum which becomes larger and narrower as one
approaches the thermodynamic limit, $N\rightarrow \infty$. In
Figure~\ref{cv_homo_perm.eps}a this is shown for the PERM model along with an
extrapolation of the CAP $\epsilon^h_c(N)$ for chains of length $N$ - see inset
- which for $N\rightarrow \infty$ becomes a straight line in agreement with
eq~\ref{Fin_epsilon}. It becomes also evident from
Figure~\ref{cv_homo_perm.eps}b that the alternative method of using the
position of the maximum of the specific heat, $C_V= (k_BT^2)^{-1} \left
(\langle U^2\rangle - \langle U \rangle ^2\right)$ from the fluctuations of the
internal energy, $U=\epsilon N_s$, does not give satisfactory results due to 
the rather flat shape of the maximum. This behavior is not surprising, if one 
recalls that the critical exponent $\alpha$ describing the divergence of $C_V$ at 
the CAP, i.e., for $\kappa \rightarrow 0$, according to  $C_V\propto \kappa^{-\alpha}$, 
see  eq~\ref{cv}, is given by $\alpha = 2 - \phi^{-1}
\approx 0$~\cite{Hsu}. It has been show
earlier\cite{Hsu}, however, that one can still use specific heat data to
determine the CAP if, instead of the position of the maximum, one examines the
common intersection point of $C_V$ vs $\epsilon$. In our simulation this yields
again $\epsilon^h_c = 0.284$ - cf. Table~\ref{table2}.

\subsubsection{From the components of $R_g$}\label{Rg_comp}
\begin{figure}[bht]
\begin{center}
\vspace{1cm}
\includegraphics[scale=0.32]{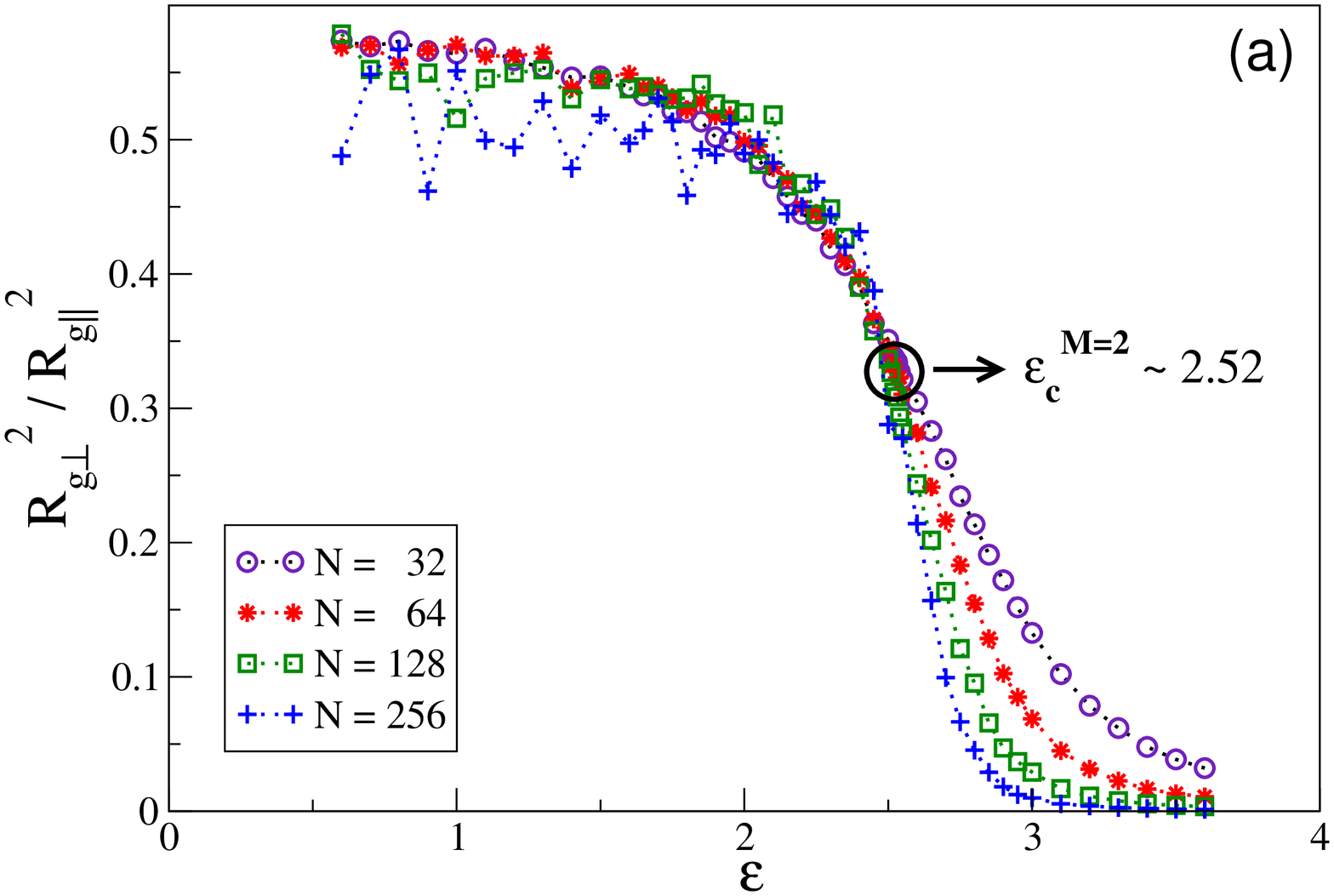}\hspace{0.7cm}
\includegraphics[scale=0.32]{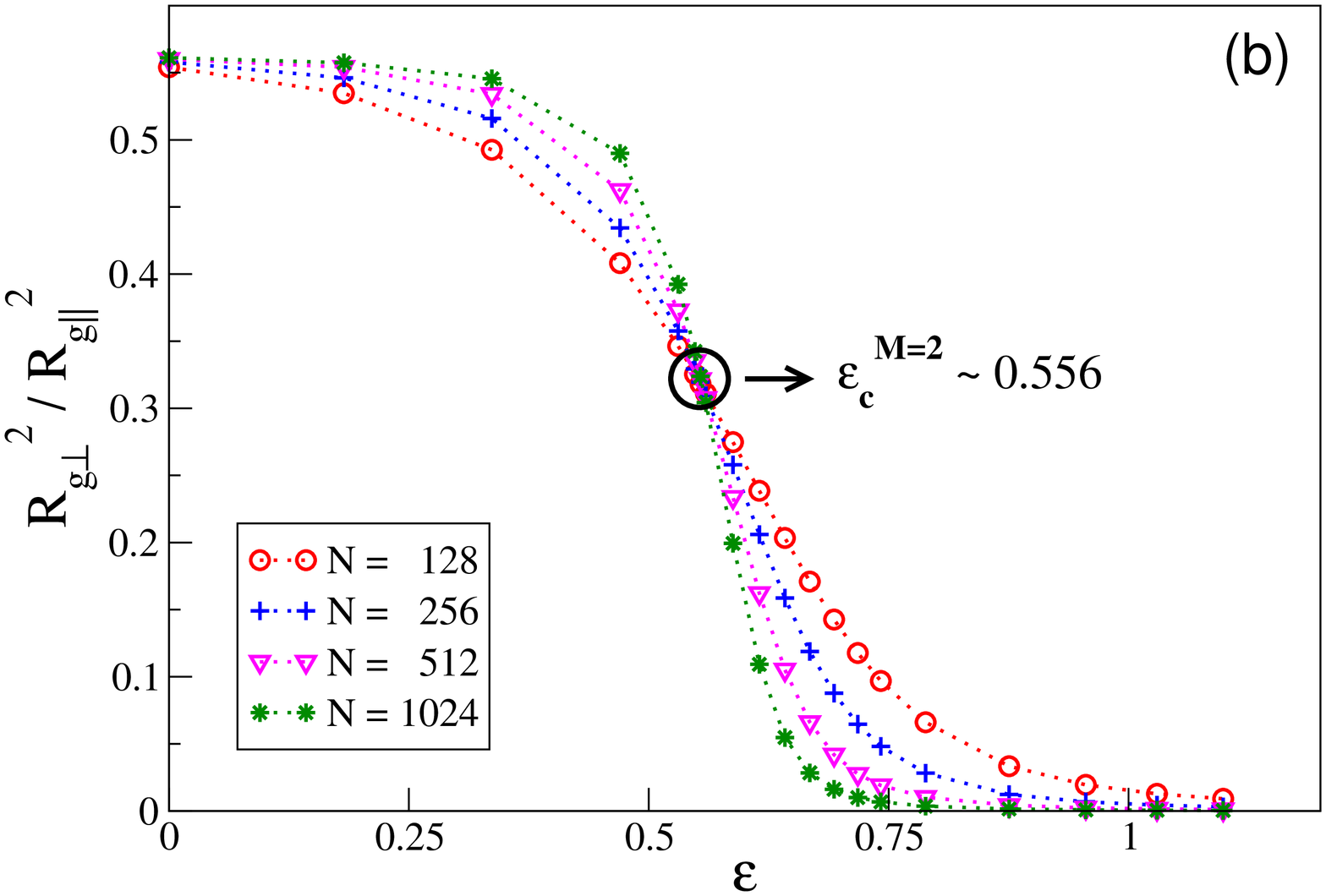}
\caption{The ratio of $R^{2}_{g\perp}/R^2_{g\parallel}$ plotted as a function of
$\epsilon$ for copolymers with block size $M=2$. The critical point is
determined by the intersection of all curves which are found to be at (a)
$\epsilon_c^{(M=2)} \approx 2.52$ (MA) and (b) $\epsilon_c^{(M=2)} \approx 0.556$ (PERM).}
\label{rg2_m2}
\end{center}
\end{figure}

According to eqs~\ref{Perp},~\ref{Parall}, and~\ref{Ratio}, one should expect
that all curves of $R_{g\perp}^2/R_{g\parallel}^2$, for different chain length
$N$ intersect at a fixed point which gives the CAP in the limit of $N\rightarrow
\infty$. In Figure~\ref{rg2_m2}, we illustrate this method by plotting the ratio
$R^2_{g\perp}/R_{g\parallel}^2$ vs $\epsilon$ for copolymers with block size
$M=2$.
For both methods, MA and PERM, the curves for different $N$ intersect nearly at 
a single intersection point, however, as before, the CAP determined by MA (see
Figure~\ref{rg2_m2}a) is less accurate than the results given by PERM (see
Figure~\ref{rg2_m2}b). The CAPs obtained from this method,
$\epsilon_c^{M=2}=2.52(3)$ by MA and $\epsilon_c^{M=2}=0.556(4)$ by PERM are
consistent with the estimates from the order parameter method where
$\epsilon_c^{M=2}=2.521(20)$ by MA and $\epsilon_c^{M=2}=0.546(8)$ by PERM. The
CAPs $\epsilon_c(M)$ for homopolymers, multi-block copolymers with different
block size $M$, and for random copolymers are listed in Table~\ref{table1}
and \ref{table2}.

\subsection{Scaling behavior}
From the data for the CAP one may check the value of the crossover exponent
$\phi = 0.50$ by plotting the order parameter $f$ vs. $N$
\begin{figure}[htb]
\begin{center}
\vspace{1cm}
\includegraphics[scale=0.32]{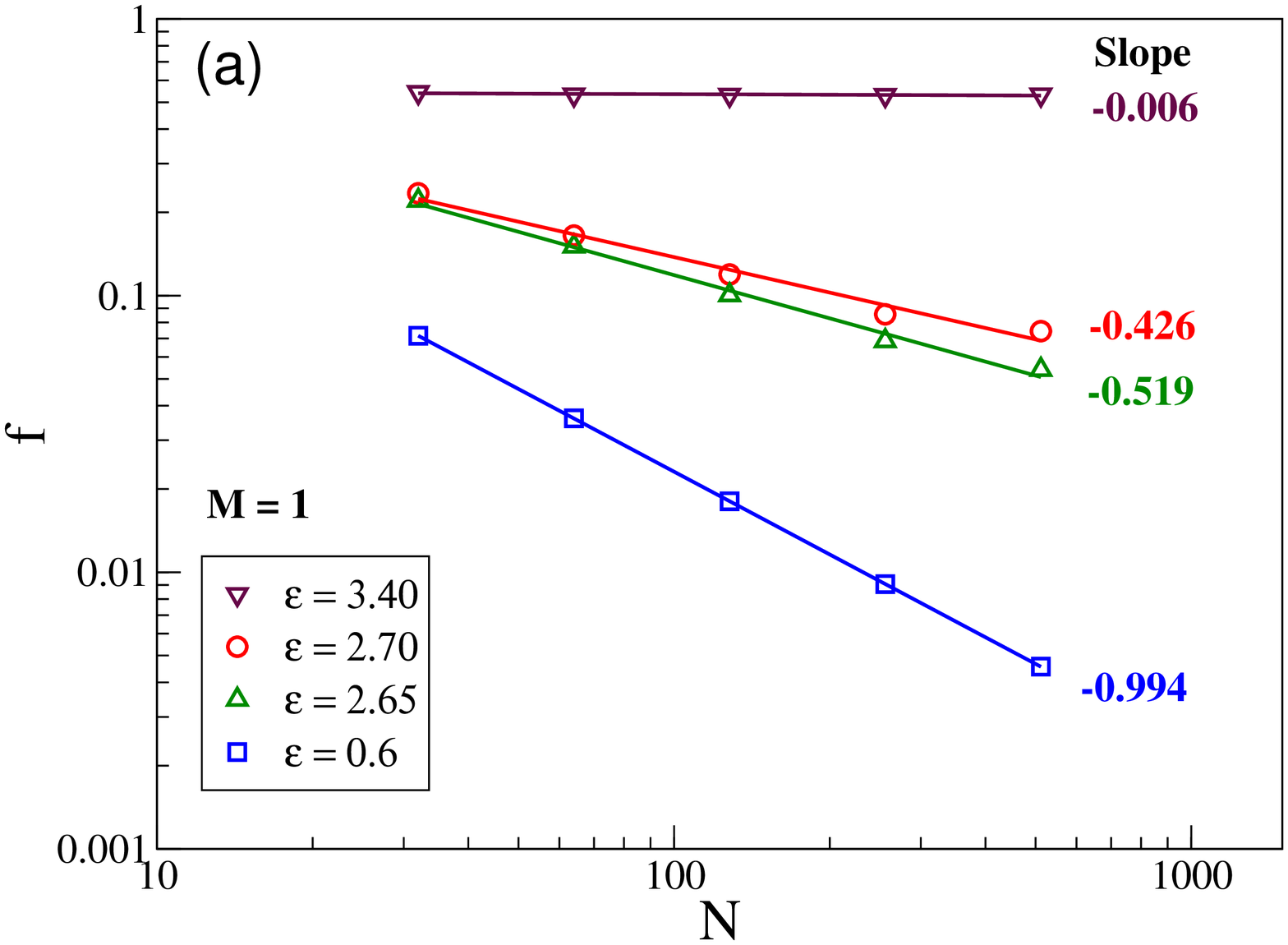}\hspace{0.7cm}
\includegraphics[scale=0.32]{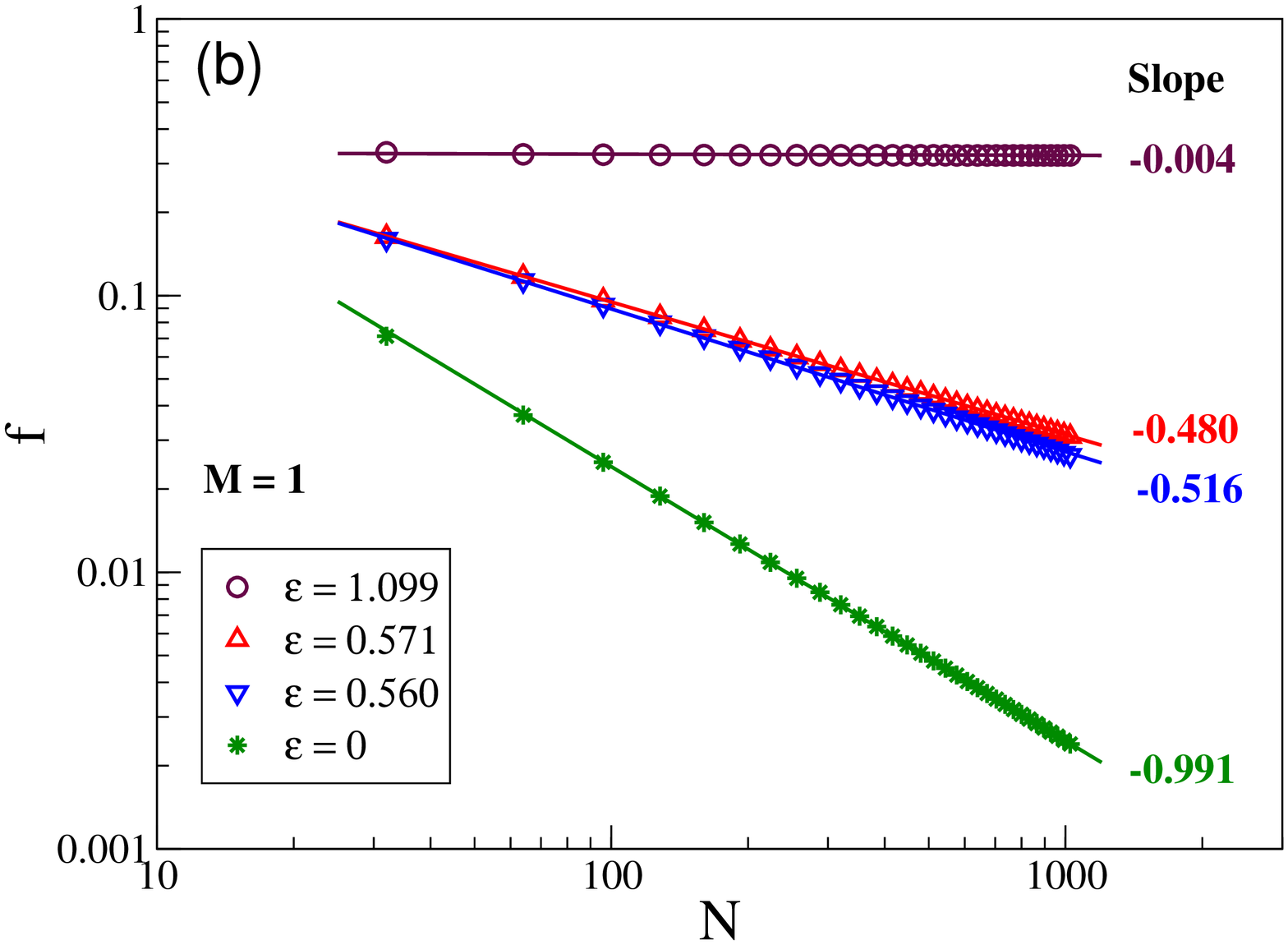}
\caption{Log-log plot of the order parameter $f$ vs N for
block copolymers with block size $(M=1)$. The value of $\epsilon$ for each
curve is given in the legend while the slope is also indicated. 
One may readily check that the straight lines with slope $0.5$ correspond 
to the respective values of $\epsilon_c$ in both models,
(a) MA and (b) PERM.}
\label{f_m1}
\end{center}
\end{figure}
(eq~\ref{Order_param}). This is illustrated in Figure~\ref{f_m1} as a double
logarithmic plot of $f$ vs. $N$ for the case of $M=1$, i.e., regular alternating
polymers. Figure~\ref{f_m1} demonstrates clearly that the slope of the $f$ vs
$N$ curves in logarithmic coordinates is equal to $1-\phi=-0.5$ only in those
cases where the strength of the substrate potential equals the CAP value
$\epsilon_c$, in agreement with the relation $f\propto N^{\phi-1}$. 
As in the case of homopolymers
(eq~\ref{Order_param}), Figure~\ref{f_m1}a shows that in the strongly adsorbed
regime ($\epsilon = 3.40$) above the CAP the order parameter $f \propto N^0$
({\em all} monomers stick). In contrast, far below the CAP, only the anchoring
monomer is attached to the substrate, $f\sim N^{-1}$, as  in the asymptotic
limit $N\rightarrow \infty$ of homopolymers. This is
observed for $\epsilon = 0.60$ for the alternating chains ($M=1$).
In Figure~\ref{f_m1}b, where the statistical precision and the chain lengths
involved are much higher, one may see that for large $N$ the curves which are
slightly above, $\epsilon=0.571$, or below, $\epsilon=0.560$, the CAP at
$\epsilon=0.568$ - cf. Figure~\ref{rg2_m2} - display slopes which differ slightly from  
$-0.5$ and thus considerably narrow the interval of critical behavor.
\begin{figure}[bht]
\begin{center}
\vspace{1cm}
\includegraphics[scale=0.32]{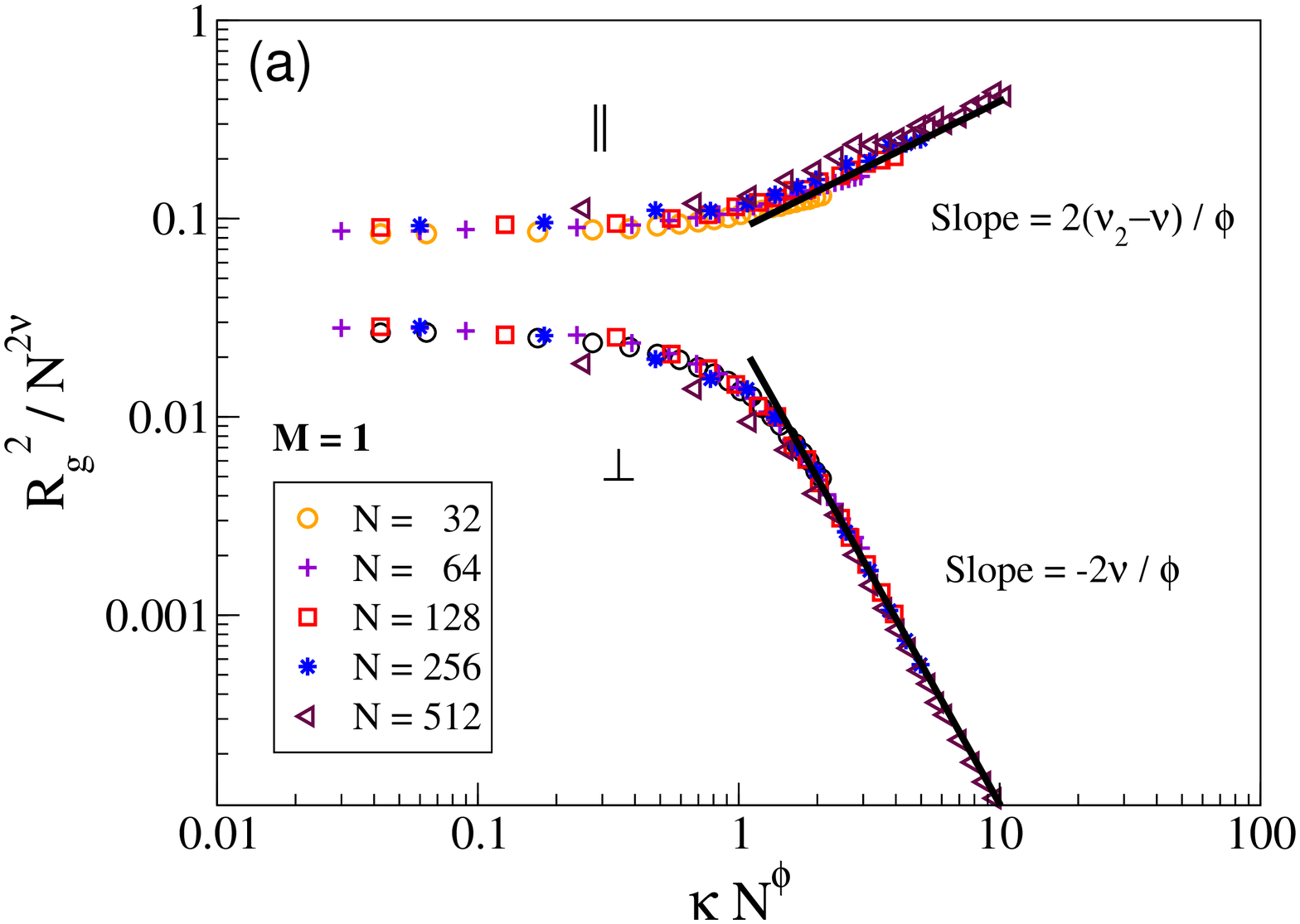}\hspace{0.7cm}
\includegraphics[scale=0.32]{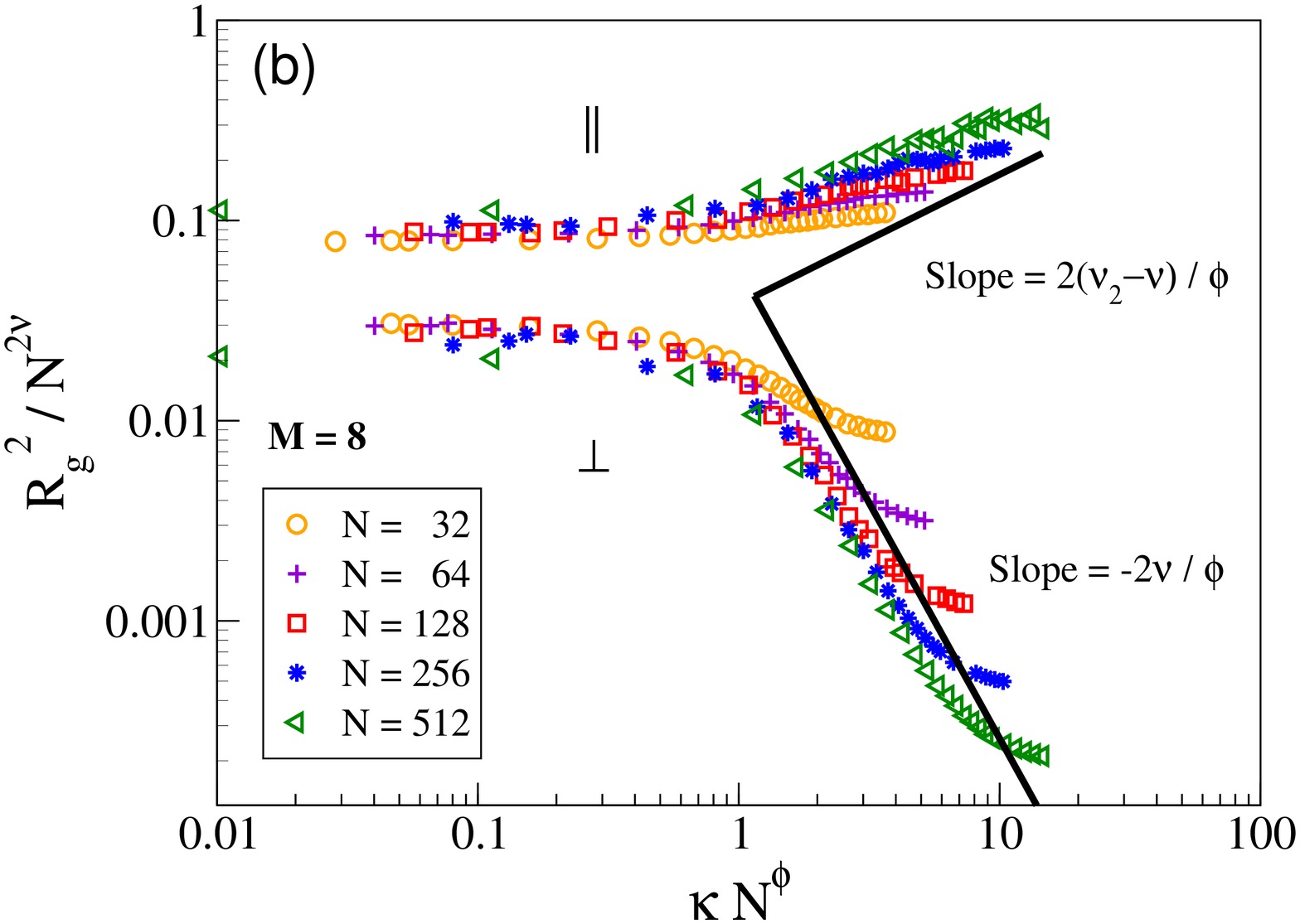}
\vskip 0.9cm
\includegraphics[scale=0.32]{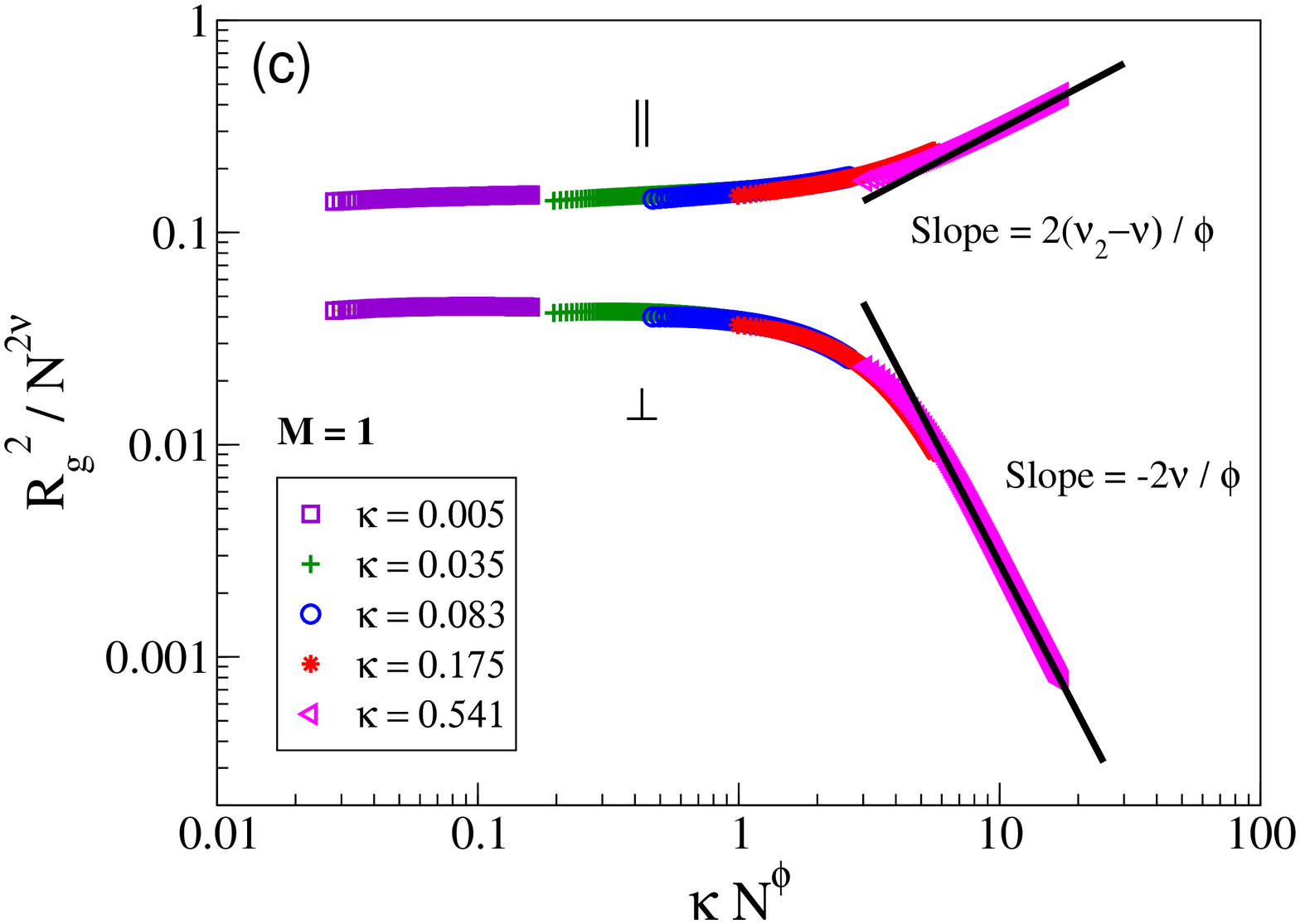}\hspace{0.7cm}
\includegraphics[scale=0.32]{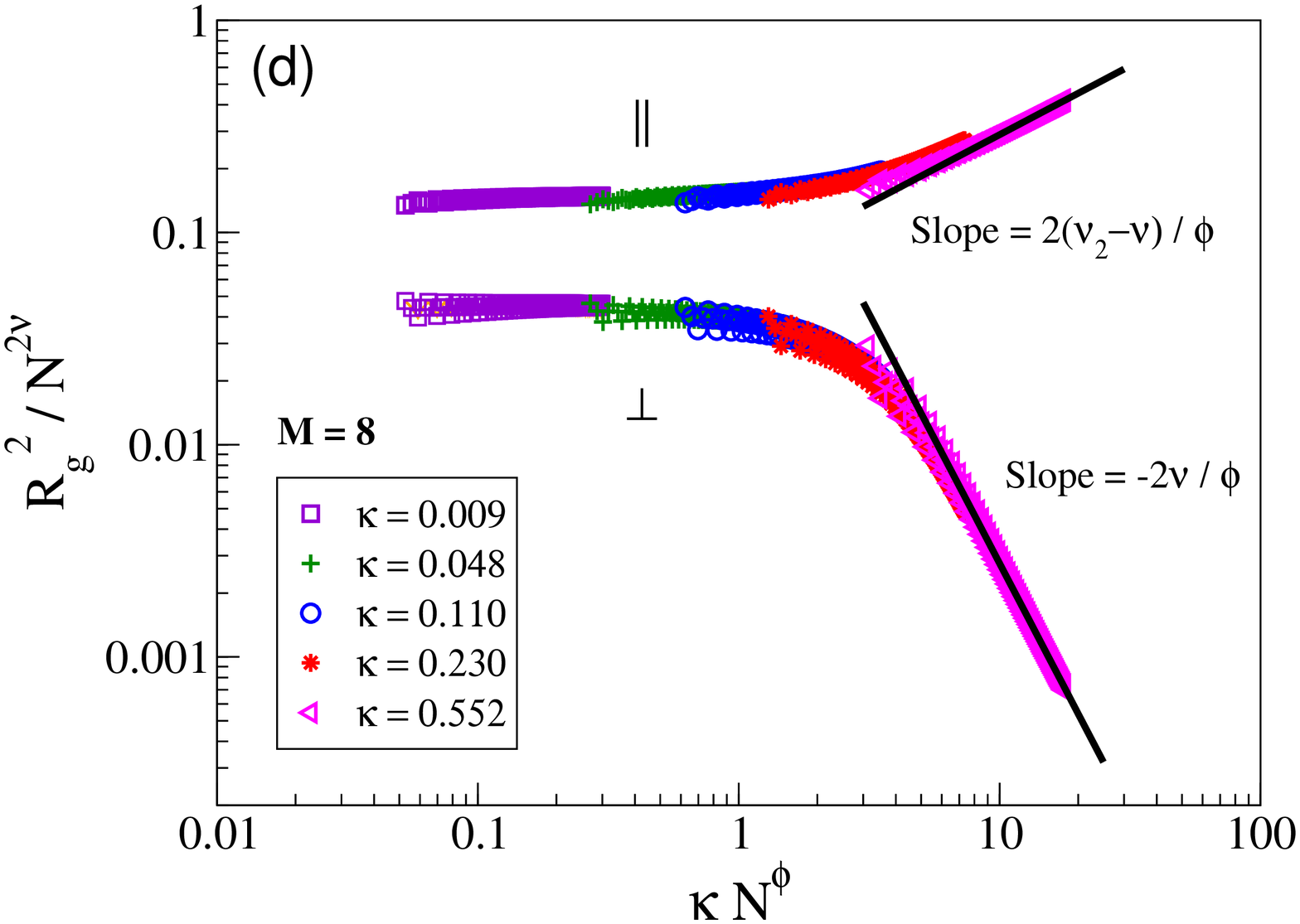}
\caption{Log-log plots of $R_{g\parallel}^2/N^{2\nu}$ and
$R^2_{g\perp}/N^{2\nu}$ vs $\kappa N^{\phi}$ with $\nu=0.588$ and $\nu_2=3/4$.
The straight lines indicate the asymptotic behaviour of the scaling functions
given by eq~\ref{Perp} and~\ref{Parall}: (a) and (b) represent results for
regular multi-block copolymers with block sizes $M=1$ and $M=8$, respectively,
and are obtained by MA; (c) and (d)  - similar results but obtained by PERM.}
\label{rgscale}
\end{center}
\end{figure}

In Figure~\ref{rgscale} we present the results for the components of the
mean square gyration radius, $R^2_{g\parallel}$ and $R^2_{g\perp}$, in 
scaled form in terms of
the parameter $\kappa N^{\phi}$ for regular block-copolymers with block size
$M=1$ and $M=8$. Generally, one observes a good agreement with the predictions
of Section~\ref{Theory}, especially concerning the data obtained by PERM - 
Figure~\ref{rgscale}c, d. 
Considerable deviations from the expected scaling behavior are observed only in
Figure~\ref{rgscale}b where the effective segment of a diblock with $M=8$ is
comparatively large for the simulated chain lengths $N\le 512$, meaning {\em
effective} chain lengths of $N_{eff} = N/16 \le 32$ which are definitely too
short for a well pronounced scaling behavior to be demonstrated. With the much
longer chains, $N\le 2048$, sampled by PERM and shown in Figure~\ref{rgscale}d,
this problem is absent. 

\subsection{Phase diagram of multi-block copolymer adsorption}

Using the values for the CAP, given in Table\ref{table1}, one may construct a
phase diagram showing the relative increase of the critical potential
$\epsilon_c(M)$ compared to that of a homopolymer against (inverse) block size
$M$. This is one of the central results of the present study.
\begin{figure}[htb]
\begin{center}
\vspace{1cm}
\includegraphics[scale=0.32]{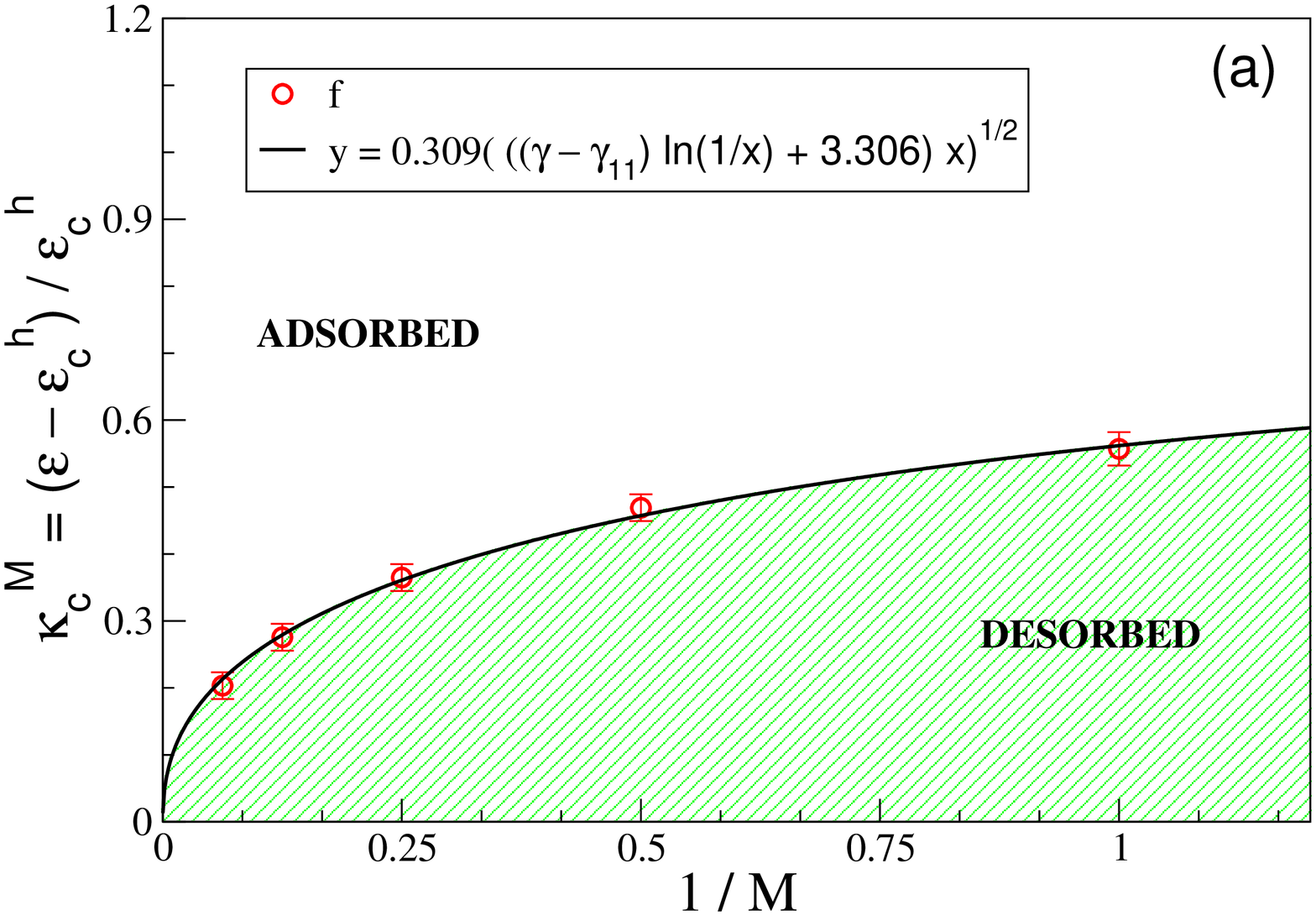}\hspace{0.7cm}
\includegraphics[scale=0.32]{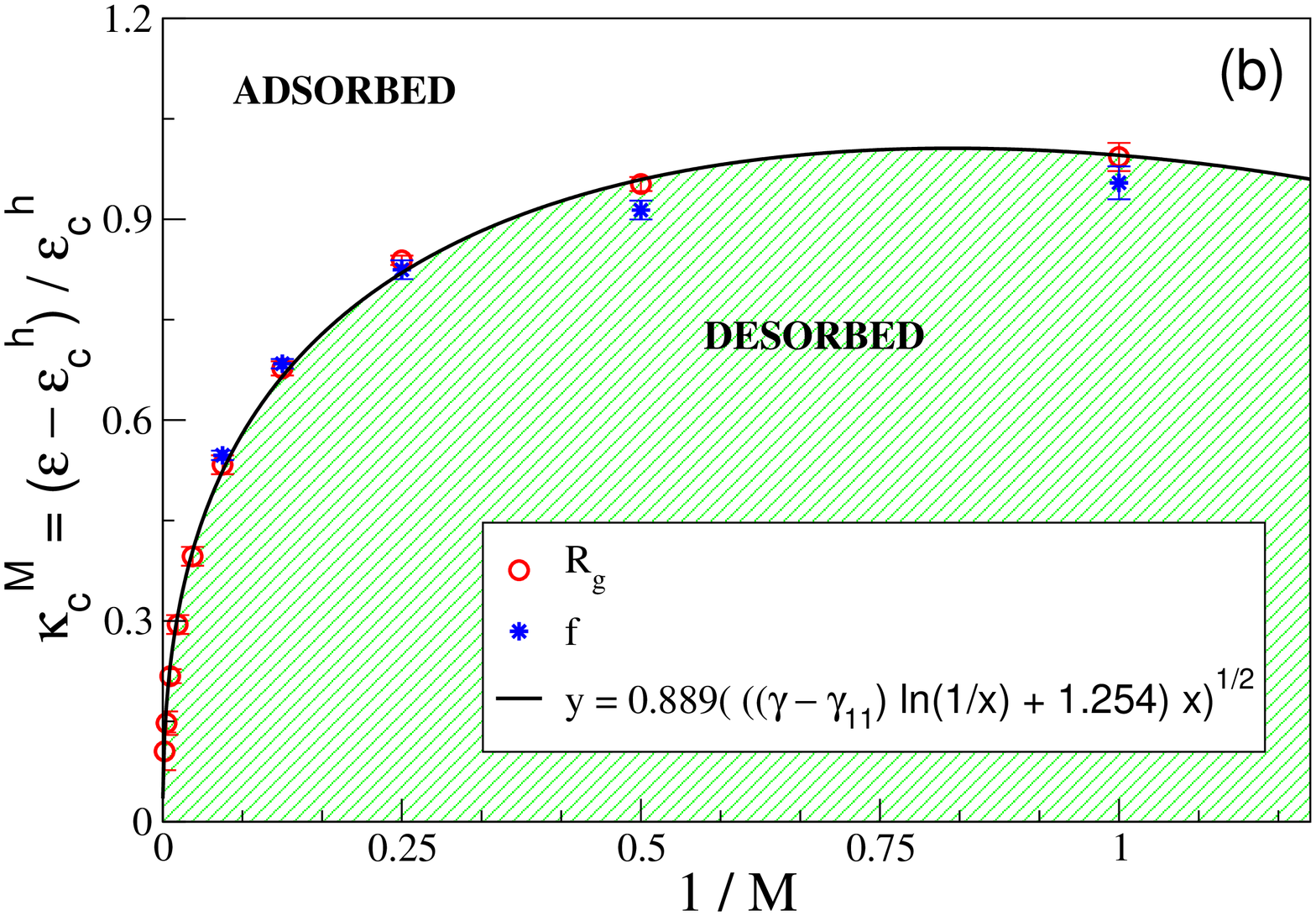}
\caption{$\kappa^M_c =(\epsilon_c^M -\epsilon_c^h)/\epsilon_c^h$
plotted vs $1/M$ for multi-block copolymers with various values of $M$.
The critical points of adsorption for homopolymers are 
(a) $\epsilon_c^h=1.716$ (MA) and (b) $\epsilon_c^h=0.285$ (PERM).
The curves give the best fit of eq~\ref{Kappa_vs_M}, $\kappa \propto
\left(\frac{(\gamma -\gamma_{11})ln(M) + E_c^h}{M} \right)^{1/2}$.}
\label{kappa_m}
\end{center}
\end{figure}
\begin{figure}[htb]
\vspace{1cm}
\includegraphics[scale=0.32]{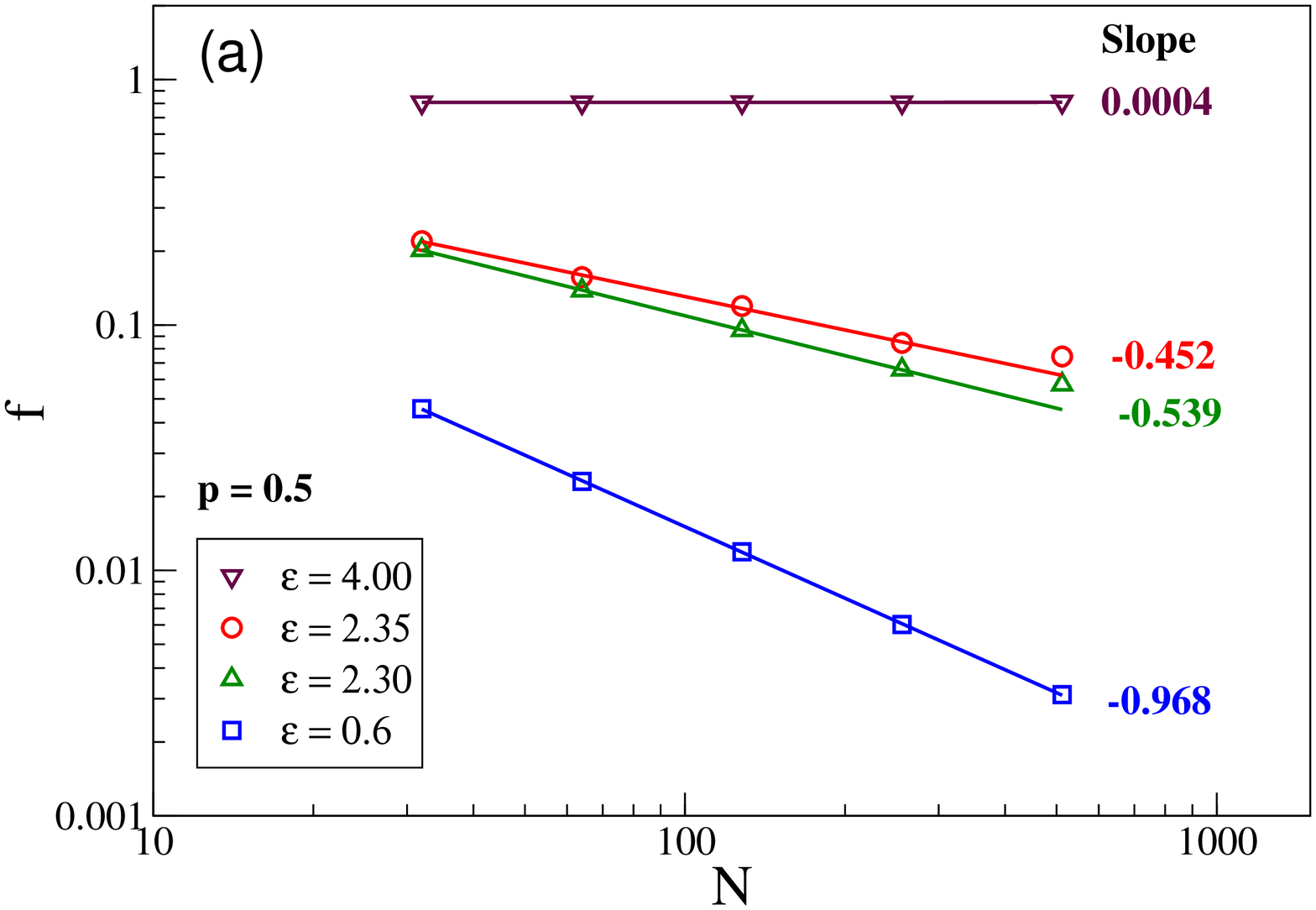}\hspace{0.7cm}
\includegraphics[scale=0.32]{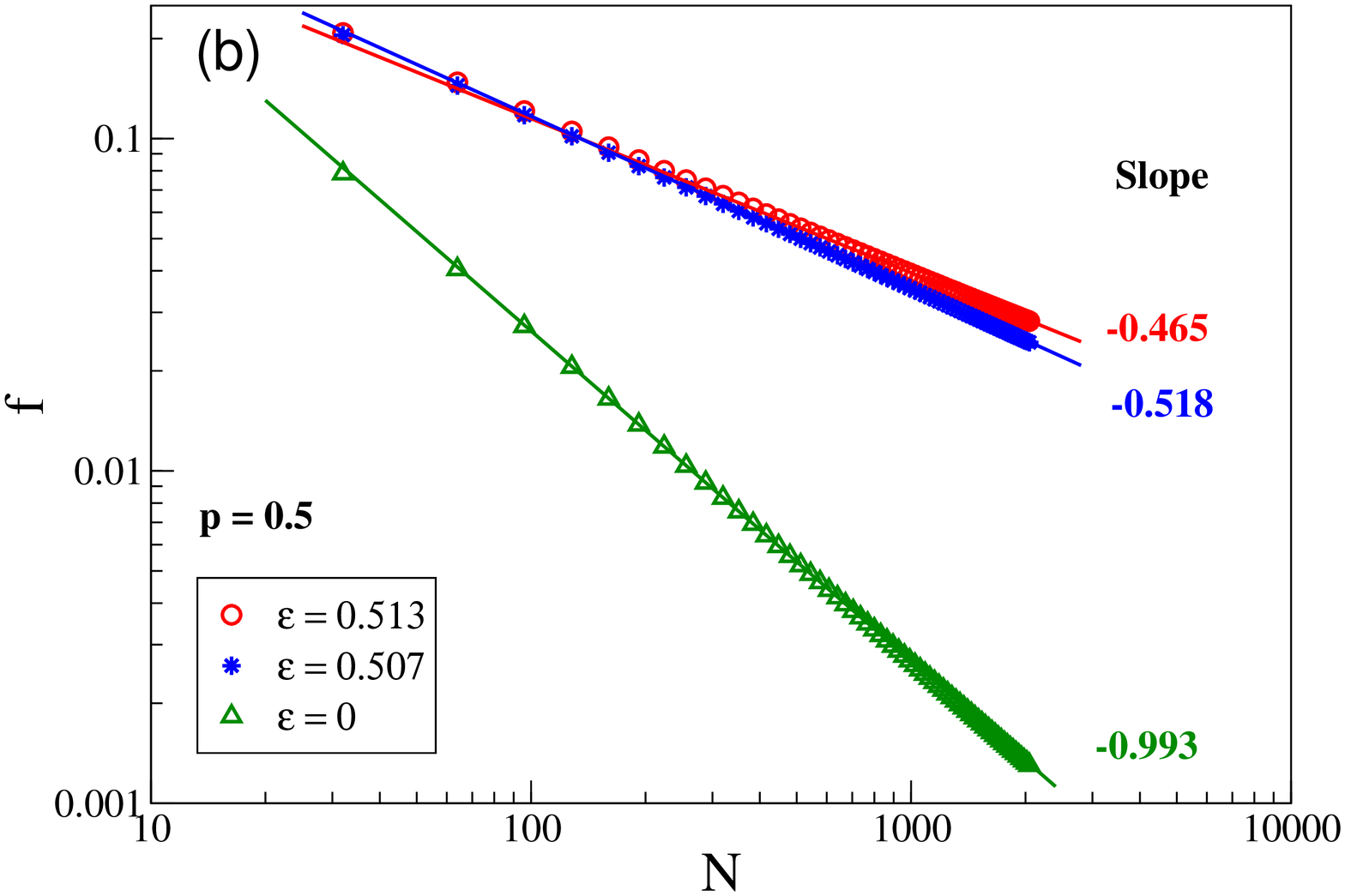}
\caption{The same as in Figure~\ref{f_m1} but for
random copolymers with the composition $p=0.5$ (a) MA where $\epsilon^p_c=2.33$
and (b) PERM with $\epsilon^p_c=0.507$.}
\label{f_p05}
\end{figure}
In Figure~\ref{kappa_m} one may see that the line of critical points, defining
the region of adsorption, for both models is a steadily growing function of the
inverse block size $M^{-1}$. Evidently, the theoretical result,
eq~\ref{Kappa_vs_M}, appears to be in good qualitative agreement with
simulation data  for the different models. As far as eq~\ref{Kappa_vs_M} comes
as a result of scaling analysis, it can be verified only up to a factor of
proportionality. As mentioned in Section~\ref{OP}, the CAP of a homopolymer,
$\epsilon^h_c$, is of the same order as that of the ``renormalized'' chain
consisting of diblocks, $E^h_c$. Thus from a fit of the data points with the
expression eq~\ref{Kappa_vs_M} one may actually determine $E^h_c$. So in the
MA model one gets $E^h = 3.306$ and for PERM $E^h = 1.254$, that is, one gets
values which are two to four times larger than the respective CAP values of a
homopolymer in both models.

\subsection{Random Copolymers}

In this section we examine the adsorption transition of random copolymers with
quenched disorder and average percentage $p$ of the $A$ monomers. In addition to
testing the scaling behavior, we also check to what extent one may employ the
theory developed within approximation of ``annealed disorder'' for the
description of the CAP properties. We performed Monte Carlo simulations for
heterogeneous random copolymers of chains lengths  $32, 64, 128, 256$ and $512$
(MA) and for $64\le N\le 2048 $ (PERM) with different fraction of attractive
monomers ($p=0.125, 0.25,0.50$ and $0.75$).

It has been pointed out earlier\cite{Baum,Moghaddam} that the crossover exponent
stays the same, $\phi=0.5$, also in the case of random copolymers. Both
simulation methods used in the present study demonstrate this in
Figure~\ref{f_p05} where qualitatively the observed picture is similar to that
of Figure~\ref{f_m1} - small deviations in the attraction potential $\epsilon$, 
which was used when sampling the values of the order parameter $f$,
manifest themselves in significant changes of the log-log slope $1-\phi$ from
the expected value of $-0.5$.
\begin{figure}[htb]
{
{\includegraphics[scale=0.32]{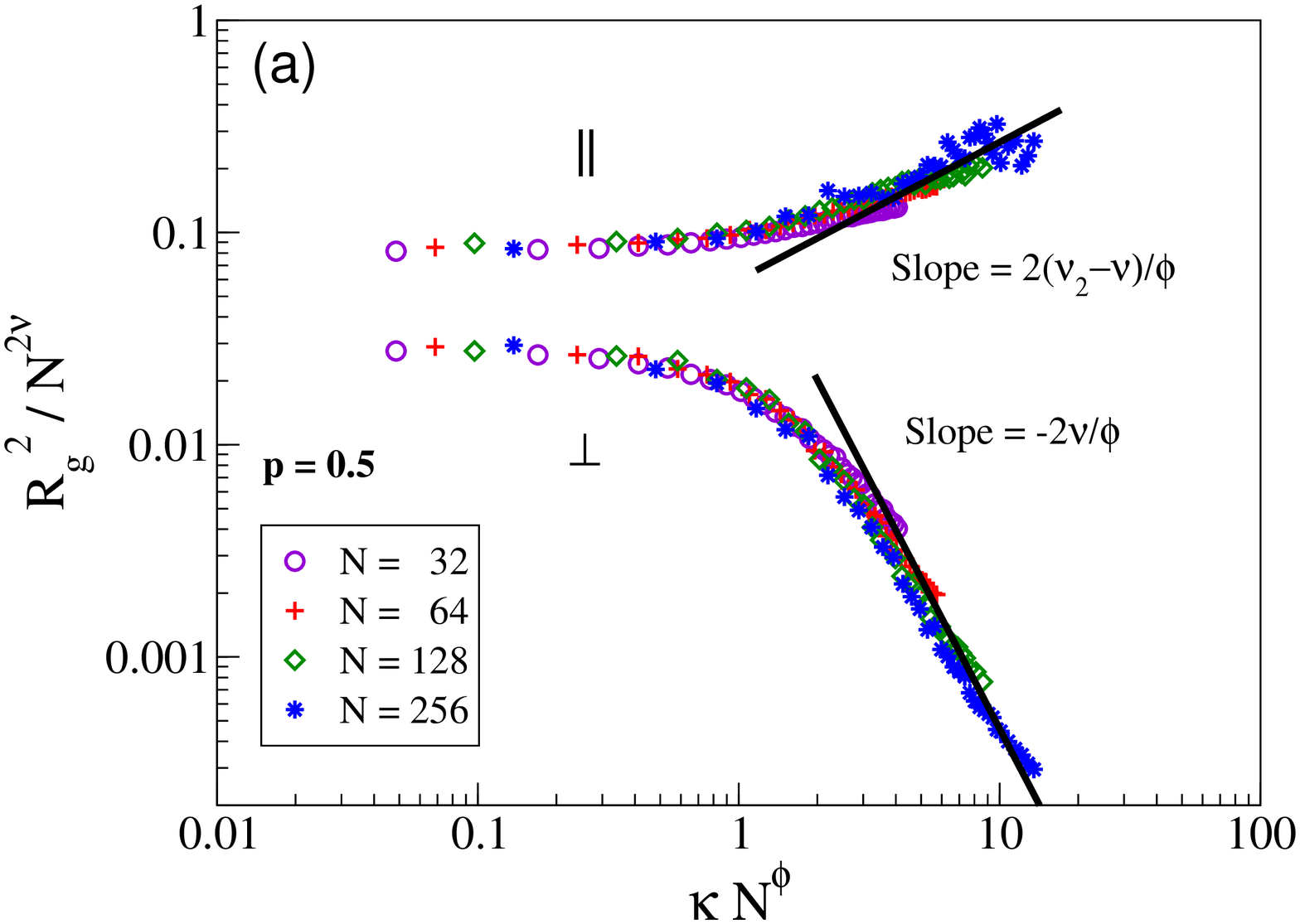}\hspace{0.7cm}
\includegraphics[scale=0.32]{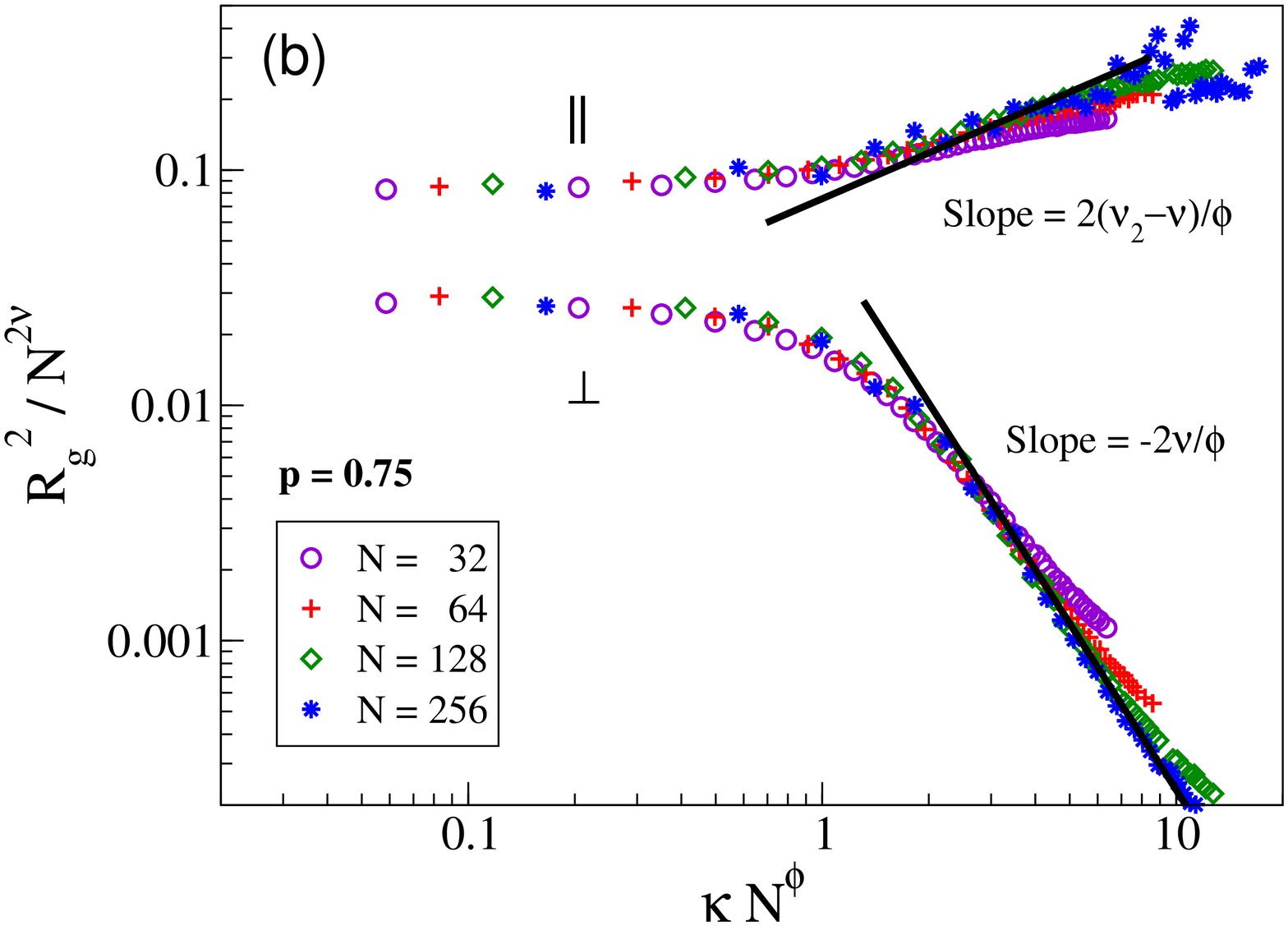} }
\vskip 1.0 true cm
{\includegraphics[scale=0.32]{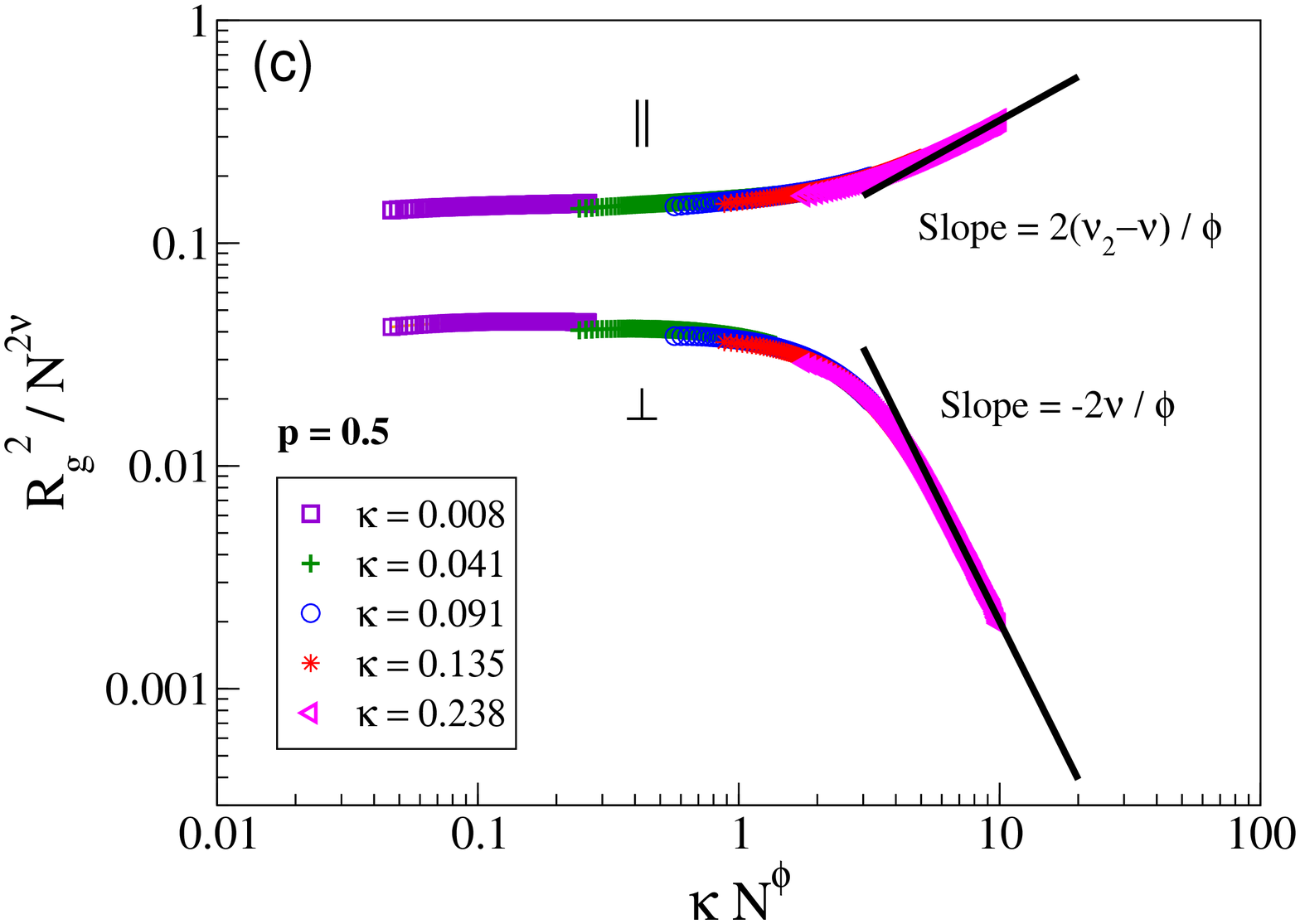}\hspace{0.7cm}
\includegraphics[scale=0.32]{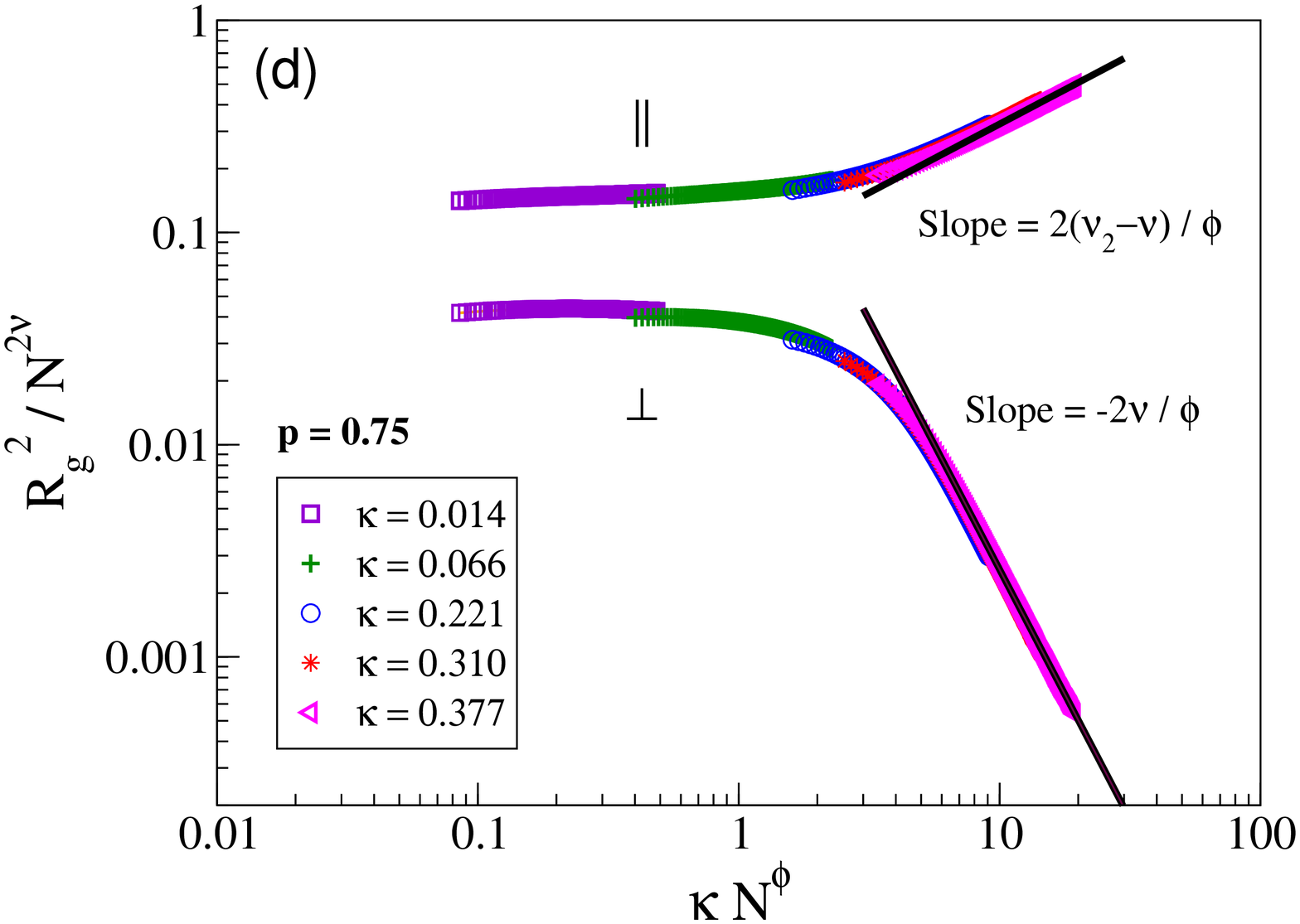}}
}
\caption{The same as in Figure~\ref{rgscale} but for random copolymers at
different composition $p$.}
\label{rgscal_p}
\end{figure}
In Figure~\ref{rgscal_p} we demonstrate that the scaling of the mena square gyration 
radius components, which we discussed before with regard to the multiblock copolymers,
holds also for random copolymers with different composition $p$. Again the
value of $\phi=0.5$ gives best scaling results. Thus it turns out that the
composition affects only the value of the CAP $\epsilon^p_c$.

\begin{figure}[thb]
\begin{center}
\vspace{1cm}
\includegraphics[scale=0.32]{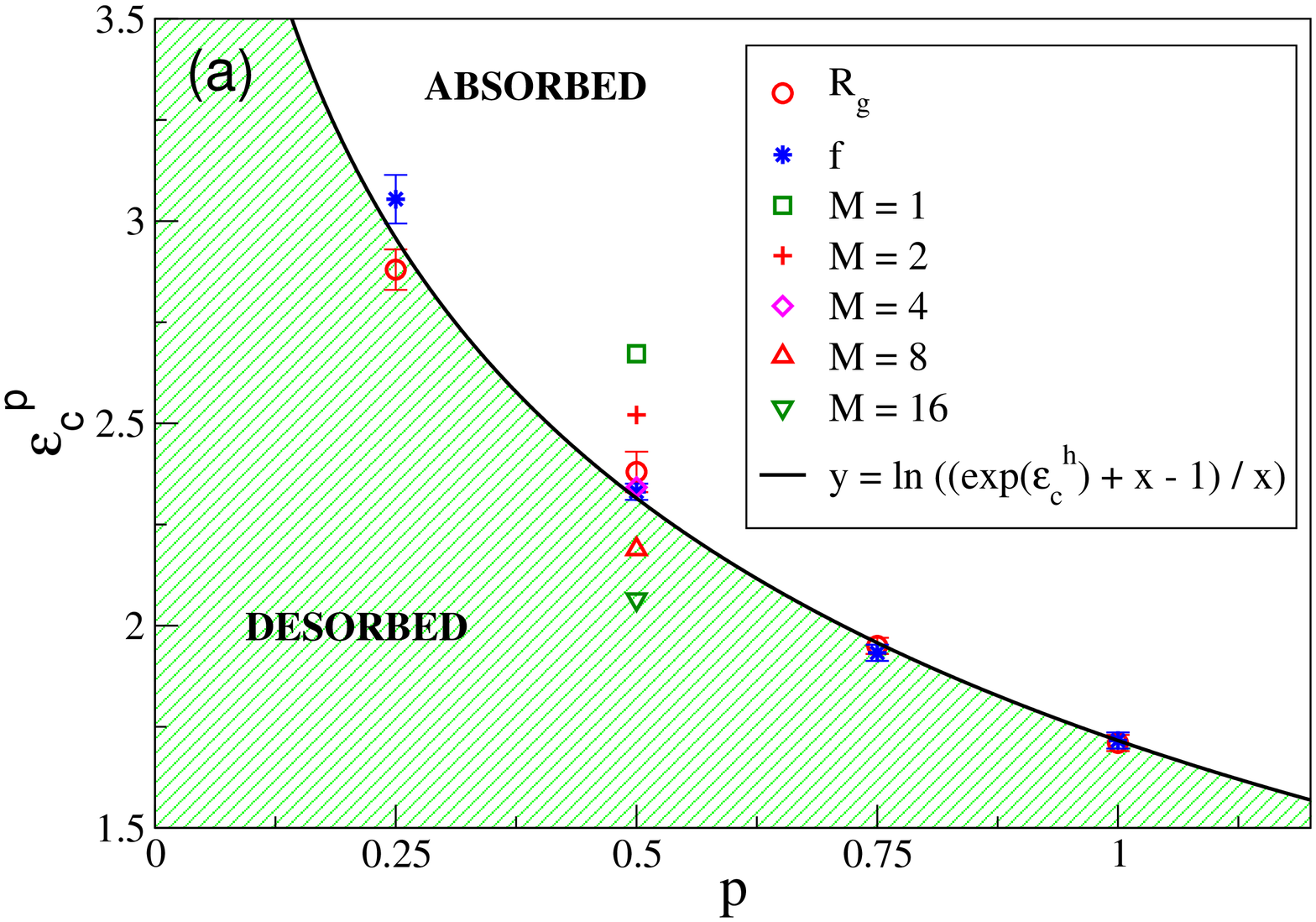}\hspace{0.7cm}
\includegraphics[scale=0.32]{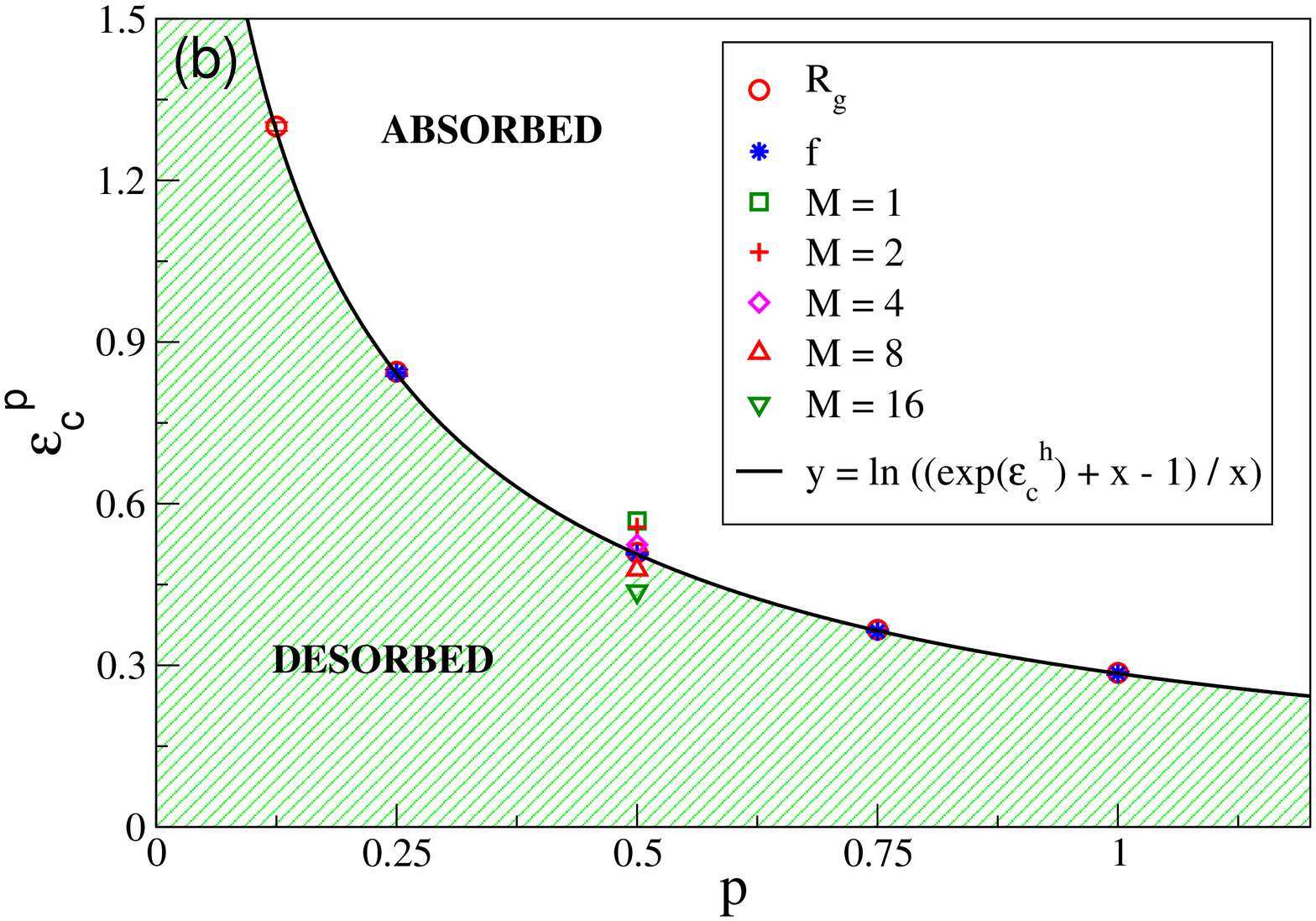}
\caption{The CAP, $\epsilon_c^p$, plotted vs the composition $p$ for random
copolymers. The curves give the best fit of eq~\ref{Copol}, $\epsilon_{c}^p =
\ln \left[ \frac{\exp{\epsilon^{h}_{c} + p - 1}}{p} \right] \geq
\epsilon^{h}_{c}$, The critical points of adsorption for homopolymers are (a)
$\epsilon_c^h=1.716$ (MA) and (b) $\epsilon_c^h=0.285$ (PERM). Symbols denote
the CAP for multiblock copolymers with block size $M$.}
\label{epsilon_p}
\end{center}
\end{figure}

In Figure~\ref{epsilon_p} we present a plot of the critical point of adsorption
against the fraction of attractive monomers. The full line corresponds to
the theoretical prediction\cite{Whit_1}, eq~\ref{Copol}. Given that there are no
fitting parameters in this equation, one finds a very good agreement between
theoretical predictions and simulation results as well as with very recent
simulation results\cite{Ziebarth} which demonstrates the the adsorption of
random copolymers can be properly described within the scope of the annealed
approximation. Figure~\ref{epsilon_p} also indicates that this approximation
breaks down for chains which are not random\cite{Ziebarth} - at $50\%$
composition the CAPs of regular block copolymers are clearly off the theoretical
prediction, eq~\ref{Copol}. As far as polymer adsorption is greatly facilitated
by the formation of trains of monomers on the substrate\cite{Ziebarth}, the
larger the block size $M$, the lower the respective CAP $\epsilon^M_c$ under
the line, eq~\ref{Copol}. No monomer trains are possible in the case of
alternating chains which results in an $\epsilon^{M=1}_c > \epsilon^p_c$. Thus
from the position of the CAPs on Figure~\ref{epsilon_p} one may conclude that
the mean length of an $A$-train on the substrate at $p=0.5$ is close to four.

\section{Concluding remarks}

The main focus of the present investigation has been aimed at the adsorption
transition of random and regular multiblock copolymers on a rigid substrate. We
have used two different models to establish an unambiguous picture of the
adsorption transition and to test scaling predictions at criticality. The
first one is an off-lattice coarse-grained bead-spring model of polymer chains
which
interact with a structureless surface by means of a contact potential, once an
$A$-monomer comes close enough to be captured by the adsorption potential. The
second one deals with SAW on a cubic lattice by the pruned-enriched Rosenbluth 
method (PERM) which is very efficient, especially for very long polymer chains, 
and provides
high accuracy of the simulation results at criticality. Notwithstanding their
basic difference, both methods suggest a consistent picture of the adsorption of
copolymers on a rigid substrate and confirm the theoretical predictions
even though the particular numeric values of the critical adsorption potential
(CAP) are model-specific and differ considerably.

As a central result of the present work, one should point out the phase diagram
of regular multiblock adsorption which gives the increase of the critical
adsorption potential $\epsilon^M_c$ with decreasing length $M$ of the adsorbing
blocks. For very large block length, $M^{-1}\rightarrow 0$, we find that the
CAP approaches systematically that of a homogeneous polymer. We demonstrate
also that the phase diagram, derived from computer experiment within the
framework of two different models, agrees well with the theoretical prediction
based on scaling considerations.

The phase diagram for random copolymers with quenched disorder which gives the
change in the critical adsorption potential, $\epsilon^p_c$, with changing
percentage of the sticking $A$-monomers, $p$, is also determined from extensive
computer simulations carried out with the two models. We observe perfect
agreement with the theoretically predicted result which has been derived by
treating the adsorption transition in terms of the ``annealed disorder''
approximation. 

We show that a consistent picture of how some basic polymer chain properties of
interest such as the gyration radius components perpendicular and parallel to
the substrate, or the fraction of adsorbed monomers at criticality, scale when
a chain undergoes an adsorption transition appears regardless of the particular
simulation approach. An important conclusion thereby concerns the value of the
universal crossover exponent $\phi=0.5$ which is found to remain unchanged,
regardless whether homo-, regular multiblock-, or random polymers are concerned.

\section{Acknowledgments}
The authors are indebted to Kurt Binder and Alexander Grosberg for discussions
of the present work. A.~M. thanks the Institute for Polymer Research, Mainz, for
hospitality during his visit. We acknowledge support from the
Deutsche Forschungsgemeinschaft (DFG), grant No. SFB 625/A3 and SFB 625/B4. 

\end{document}